\documentclass[aps,prd,a4paper,reprint,nofootinbib,superscriptaddress,floatfix,preprintnumbers]{revtex4-2}
\usepackage[dvipsnames]{xcolor}
\usepackage{xurl}
\usepackage{hyperref}
\usepackage{amsmath}
\usepackage{amsfonts}
\usepackage{amssymb}
\usepackage{color}
\usepackage[utf8]{inputenc}
\usepackage{graphicx}
\usepackage{array}
\usepackage{microtype}
\usepackage{siunitx}
\usepackage{soul}
\usepackage{longtable}
\usepackage{bm}
\usepackage{xspace}
\usepackage{natbib}
\usepackage[normalem]{ulem}
\usepackage[nameinlink,capitalise]{cleveref}

\usepackage{snapshot} 

\newcommand{\de}{{\rm d}}

\newcommand{\be}{\begin{equation}}
\newcommand{\ee}{\end{equation}}

\newcommand{\bff}{\begin{figure}[t!]}
\newcommand{\eff}{\end{figure}}

\let\Gamma\varGamma
\let\Delta\varDelta
\let\Theta\varTheta
\let\Lambda\varLambda
\let\Xi\varXi
\let\Pi\varPi
\let\Sigma\varSigma
\let\Upsilon\varUpsilon
\let\Phi\varPhi
\let\Psi\varPsi
\let\Omega\varOmega

\begin{document}

\title{Dark Matter and Galaxy Cross-Correlations \\ with the Cherenkov Telescope Array Observatory} 

\date{\today}
\author{Elena Pinetti}\email{epinetti@flatironinstitute.org}
\affiliation{Center for Computational Astrophysics - Cosmology, Flatiron Institute, New York, NY 10010, USA}
\affiliation{Emmy Noether Fellow, Perimeter Institute for Theoretical Physics, 31 Caroline Street N., Waterloo, Ontario N2L 2Y5, Canada}
\affiliation{Fermi National Accelerator Laboratory, Theoretical Astrophysics Department, Batavia IL, 60510, USA}
\affiliation{University of Chicago, Kavli Institute for Cosmological Physics, Chicago IL 60637, USA}
\author{Veronika Vodeb}\email{veronika.vodeb@ung.si}
\affiliation{University of Nova Gorica, Center for Astrophysics and Cosmology}
\author{Aurelio Amerio}\email{aurelio.amerio@ific.uv.es}
\affiliation{Instituto de F\'isica Corpuscular (IFIC), University of Valencia and CSIC, Calle Catedrático José Beltrán 2, 46980 Paterna, Spain}
\author{Alessandro Cuoco}\email{alessandro.cuoco@unito.it}
\author{Stefano Camera}\email{stefano.camera@unito.it}
\affiliation{Dipartimento di Fisica, Universit\`a degli Studi di Torino, via P.\ Giuria, 1 10125 Torino, Italy}
\affiliation{INFN -- Istituto Nazionale di Fisica Nucleare, Sezione di Torino, via P.\ Giuria 1, 10125 Torino, Italy}
\affiliation{INAF -- Istituto Nazionale di Astrofisica, Osservatorio Astrofisico di Torino, strada Osservatorio 20, 10025 Pino Torinese, Italy}
\author{Nicolao Fornengo}\email{nicolao.fornengo@unito.it}
\affiliation{Dipartimento di Fisica, Universit\`a degli Studi di Torino, via P.\ Giuria, 1 10125 Torino, Italy}
\affiliation{INFN -- Istituto Nazionale di Fisica Nucleare, Sezione di Torino, via P.\ Giuria 1, 10125 Torino, Italy}
\author{Gabrijela Zaharijas}\email{gabrijela.zaharijas@ung.si}
\affiliation{University of Nova Gorica, Center for Astrophysics and Cosmology}

\begin{abstract}
The Cherenkov Telescope Array Observatory (CTAO) will be a ground-based Cherenkov telescope performing wide-sky surveys, ideal for anisotropy studies such as cross-correlations with tracers of the cosmic large-scale structure. Cross-correlations can shed light on high-energy $\gamma$-ray sources and potentially reveal exotic signals from particle dark matter. In this work, we investigate CTAO sensitivity to cross-correlation signals between $\gamma$-ray emission and galaxy distributions. We find that by using dense,  low-redshift catalogs like 2MASS, and for integration times around 50 hours, this technique achieves sensitivities to both annihilating and decaying dark matter signals that are competitive with those from dwarf galaxy and cluster analyses.
\end{abstract}

\maketitle

\section{Introduction}\label{sec:Intro}

{High-energy cosmic messengers offer a glimpse into some of the most powerful phenomena in the Universe. Among these messengers, $\gamma$-rays stand out as a unique tool for exploring the physical processes driving astrophysical sources such as active galactic nuclei (AGN), supernovae, star-forming galaxies, and pulsars. At the same time, $\gamma$-rays are among the most promising observational signatures expected from dark matter (DM), should it consist of a new particle capable of self-annihilation or decay.}

The Cherenkov Telescope Array Observatory (CTAO) \cite{CTAConsortium:2018tzg, mazin2019cherenkov}, designed to detect cosmic $\gamma$-rays with energies above 20 GeV, will offer a unique opportunity to probe both the high-energy emission of astrophysical sources and potential signals from DM particles with masses in the TeV range. CTAO will improve upon the performance of current imaging atmospheric Cherenkov telescopes, thanks to its better angular resolution, larger field of view, and broader energy coverage. It will also be the first ground-based Cherenkov telescope to carry out an intensive program of dedicated sky surveys {\cite{CTAConsortium:2018tzg, mazin2019cherenkov}}, covering large regions (i.e.\ more than a quarter) of the sky. These capabilities make CTAO particularly well-suited to studying the cosmological $\gamma$-ray emission potentially produced by DM annihilation or decay. However, this emission is irreducibly intermingled with the non-thermal emissions from astrophysical sources. The diffuse unresolved $\gamma$-ray background \cite{Fornasa_2015} (UGRB) consists of the cumulative emission from $\gamma$-ray astrophysical sources which are too faint to be individually resolved, with a possible additional contribution from DM \cite{Ajello:2015mfa}.

A potentially powerful method to separate the astrophysical emission from a DM contribution is the analysis of anisotropies in the $\gamma$-ray sky.
Various techniques are possible for this purpose. 
The present work focuses in particular on the auto-correlation of the gamma-ray signal and 
cross-correlations between the $\gamma$-ray emission and a gravitational tracer of the DM distribution in the Universe.
Auto-correlation analysis are especially well suited for studying the properties of rare and bright astrophysical populations, like blazars, as opposed to faint and numerous populations, such as star-forming galaxies \cite{Ando:2006mt,Ando:2013ff,Fermi-LAT:2018udj,fornasa2016angular,Korsmeier:2022cwp}.
Anisotropy auto-correlation studies specifically in the TeV band for Cherenkov telescopes have been performed in Refs. \cite{Hutten:2018wop,Hutten:2016jko,Ripken:2012db}.
The cross-correlation technique, instead, offers the advantage of incorporating direct gravitational tracers of the matter distribution in the Universe, and can leverage differences in angular scale, energy spectrum, and redshift evolution between the cross-correlated fields \cite{camera2013novel,camera2015tomographic,fornengo2014particle}. This approach provides a multi-dimensional handle to disentangle signals with distinct features in their spatial distribution, energy spectra, and redshift dependence. Extragalactic astrophysical sources, for instance, typically appear as point-like emitters at $\gamma$-ray energies, whereas DM is expected to produce a diffuse signal that traces the large-scale structure of the Universe. Different classes of astrophysical sources exhibit diverse spectral energy distributions, which are generally distinct from those predicted for DM-induced emissions. Moreover, their redshift distributions differ: DM $\gamma$-ray emission is expected to peak at low redshift, unlike unresolved astrophysical populations {\cite{camera2013novel,camera2015tomographic,fornengo2014particle}}. As a result, DM signals are expected to correlate more strongly with low-redshift gravitational tracers.
Beyond its application in the search for DM, the cross-correlation technique also holds broader potential for studying unresolved $\gamma$-ray source populations, offering insights into their redshift distribution and clustering properties.

Cross-correlation analyses have been performed between {\it Fermi}-LAT $\gamma$-ray sky maps and various tracers of the large-scale structure. These include weak gravitational lensing  \cite{shirasaki2014cross,shirasaki2016cosmological,troster2017cross,DES:2019ucp}, the clustering of galaxies \cite{xia2015tomography,regis2015particle,cuoco2015dark,shirasaki2015cross,cuoco2017tomographic,Ammazzalorso:2018evf,paopiamsap2024constraints}, galaxy clusters \cite{branchini2017cross,shirasaki2018correlation,hashimoto2019measurement,colavincenzo2020searching,tan2020bounds}, and the lensing effect of the cosmic microwave background \cite{fornengo2015evidence}, which traces %
the large-scale distribution of matter across cosmological distances. Additional studies are presented in Refs. \cite{ando2014mapping,fornengo2014particle,fornasa2016angular,feng2017planck,Tan:2020fbc}). 
Positive cross-correlation signals have been found with galaxies at the level of $3.5\sigma$ \cite{xia2015tomography,regis2015particle,cuoco2015dark} up to $8-10\sigma$ \cite{paopiamsap2024constraints}, galaxy clusters at $4.7\sigma$ \cite{branchini2017cross}, lensing of the cosmic microwave background at $3.2\sigma$ \cite{fornengo2015evidence} and weak gravitational lensing at $5.3\sigma$ \cite{DES:2019ucp} and more recently at signal-to-noise ratio of 8.9  \cite{DES:2025ulp}.

In this work, we explore the potential of the cross-correlation and auto-correlation techniques to detect a signal originating from sub-threshold extragalactic source populations or from annihilation or decay of TeV-scale DM particles by using the CTAO extragalactic survey capabilities.

\section{The Cherenkov Telescope Array}\label{sec:CTA}
The CTAO will represent the next generation of imaging atmospheric Cherenkov telescopes (IACTs), improving the technology of the current ground-based $\gamma$-ray detectors, such as H.E.S.S., MAGIC, and VERITAS. It is designed to surpass the sensitivity of current IACTs by an order of magnitude in the TeV regime, to improve the angular resolution, as well as to enhance the surveying and monitoring capabilities. CTAO  will consist of two IACT arrays: one located in the Southern hemisphere (CTAO-South) and the other in the Northern hemisphere (CTAO-North) \cite{CTAConsortium:2018tzg}.

The configuration will use three different types of telescope: large-, medium-, and small-sized telescopes (LST, MST, and SST, respectively), detecting photons over a broad energy range from 20 GeV up to 300 TeV. The LSTs will be most sensitive at lower energies, while the SSTs will cover the highest energies. The SSTs will be {dedicated primarily to the observation of}  phenomena occurring within our Galaxy and will {therefore be deployed only at the southern site from which} the central/inner part of the Galaxy is most visible. LSTs and MSTs will instead be built both in the Northern and Southern locations and will focus on observations of extragalactic targets.

The telescopes will be equipped with cameras featuring wide fields of view: 
{4.3$^\circ$} for LSTs, {7.5--7.7$^\circ$} for MSTs, and {8.8$^\circ$} for SSTs. A wide field of view is important for observing extended sources and diffuse emission, as well as for reducing the systematic errors of the measurements when requiring a uniform response over regions much larger than the telescope point-spread function (PSF). Several surveys covering large portions of the sky are planned as part of CTAO's observational strategy, and will be performed as Key Science Projects (KSPs) of the observatory early in the observatory's operation 
\cite{CTAConsortium:2018tzg}. Thanks to its improved sensitivity and wider field of view, {CTAO will surpass} the surveying capabilities of existing IACTs. In particular, the Extra-Galactic survey (detailed below) { will be} the largest survey ever undertaken by IACTs.   

At the same time, CTAO will carry out a large number of traditional, dedicated pointings across the sky. The off-source data collected through these observations will cumulatively cover a substantial sky area with a significant average exposure. In this work, we consider extragalactic data obtained from both observational strategies, as detailed in two subsections below.

\subsection{The Extra-Galactic survey}

To study the auto- and cross-correlation signals with gamma-rays, {it is essential to use maps of large sky regions with approximately uniform coverage}. Among CTAO Key Science Projects, the dedicated survey of the extragalactic sky---referred to as the extra-galactic (EGAL) survey - presents the required characteristics.

As described in \cite{CTAConsortium:2018tzg} (Sect. 8), the EGAL survey will cover a quarter of the sky, targeting the region defined by Galactic coordinates $-90^\circ < l < 90^\circ$ and $b > 5^\circ$. The survey is planned to reach an integral sensitivity $\sim$ 6 mCrab above 125 GeV. One of its main objectives is to provide a high-resolution map of the extragalactic sky for photons with energies at least up to 10 TeV and to compile a catalog of very high energy (VHE) extragalactic sources.

The EGAL observations are planned to be carried out over three years of observations, using the full CTAO Observatory (both the Southern and Northern arrays) to achieve nearly uniform sensitivity over the survey region. It is proposed that 15\% of the sky will be covered by the Southern array using 400 h and 10\% by the Northern array using 600 h, resulting in a total observation time of $\sim$ 1000 h. The default {observing strategy consists of a grid spaced by $\sim 3^\circ$ across the survey area, with each pointing observed for 0.51 h in the South and 1.11 h in the North. This configuration is expected to yield an effective exposure of approximately 3 h per point within the EGAL survey boundary.} 

\subsection{Simulating survey observations}
To simulate the planned EGAL survey observations, we use the software package \texttt{ctools},\footnote{\url{http://CTA.irap.omp.eu/ctools/}.} developed for a scientific analysis of CTA data. 
We used the instrument response functions (IRFs) for the observatory,\footnote{See \url{http://www.CTA-observatory.org/science/CTA-performance/} for more details.} labeled \texttt{prod3b-v2}, defining the so-called ``omega'' configuration. The respective IRF data files are publicly available at Ref. \cite{zenodo}.

\begin{figure}[t!]
    \centering
    \includegraphics[width=0.45\textwidth]{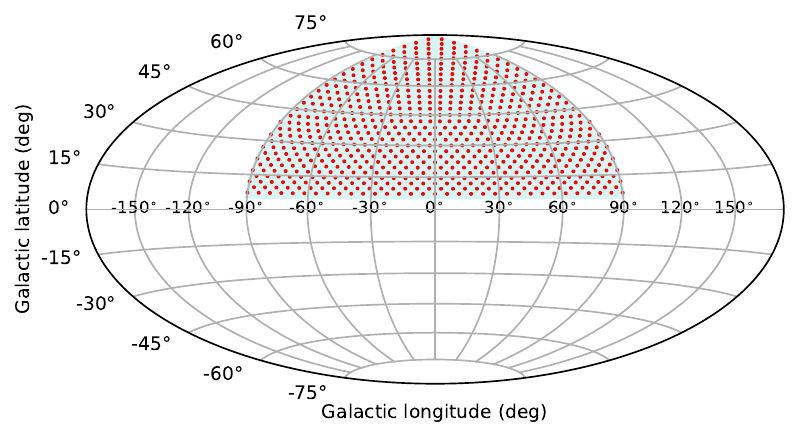}
    \caption{Pointing directions for the region of the sky {planned to be covered by the extragalactic survey. The pointing directions are shown using a HEALPix pixelization with NSIDE = 16.}}
    \label{fig:pnts}
\end{figure}

To enclose the region of the sky covered by the EGAL survey, we simulate observations using telescope-pointing directions corresponding to a HealPix realization with \texttt{NSIDE} = 16, resulting in 694 (3072) pointings within the EGAL survey bounds (over the whole sky). With this setup, the typical angular separation between adjacent pointings is $\sim 3.7^\circ$. {For definiteness,} the radius of the field of view (FoV) of each pointing is set to $6^\circ$. The pointing directions used for the simulation of the EGAL survey within the region of interest (ROI) are shown in \cref{fig:pnts}.

\begin{figure}[t!]
    \centering
    \includegraphics[width=0.45\textwidth]{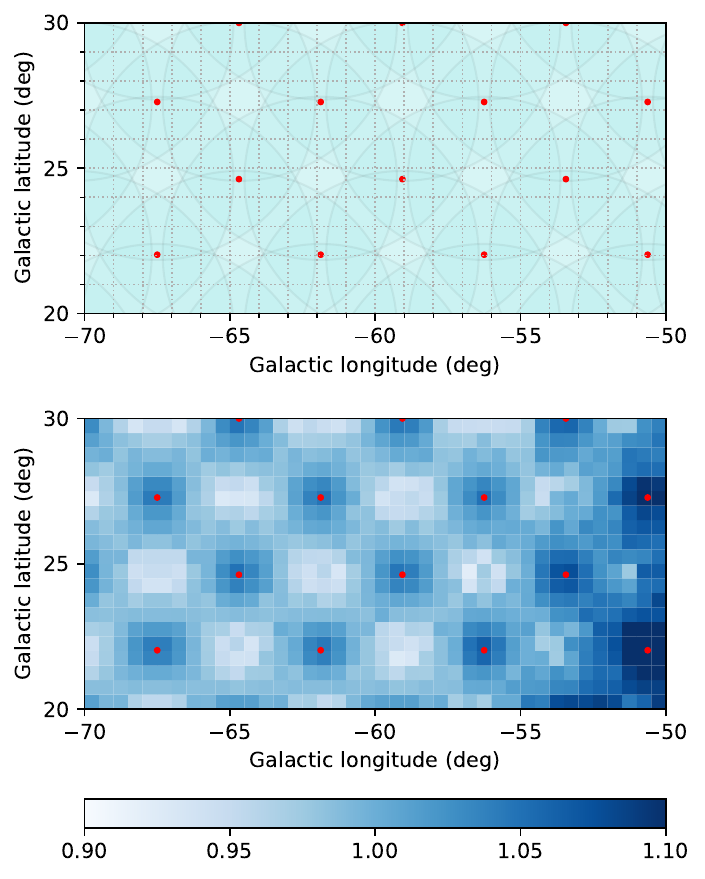}
    \caption{Overlapping areas of different pointings. \emph{Top:} A zoomed-in scheme of the pointing strategy shown in \cref{fig:pnts} (red), {together with the overlapping fields of view of individual pointings} (turquoise). \emph{Bottom:} {Simulated exposure map showing variations relative to the average exposure, highlighting the anisotropic pattern created by overlapping pointing regions.}}
    \label{fig:pnt_zoom}
\end{figure}

The overlap of the FoV from different pointings contributes to a more uniform coverage of the survey’s ROI, minimizing the fluctuations in the exposure that inevitably occur when smaller FoV observations are used to cover a large survey area. Depending on the selected \texttt{HealPix} grid size (i.e., the values of the \texttt{NSIDE} parameter) and the pointing radius for individual observations, the overlap between nearby pointings can vary in extent. The overlaps between pointings are schematically shown in \cref{fig:pnt_zoom}, where the upper panel presents a zoomed-in scheme of the pointing strategy from \cref{fig:pnts}, and the bottom panel shows the simulated exposure of the same region. These images demonstrate that the overlap of nearby pointings creates an anisotropy pattern in the corresponding observation exposure, with variations of approximately $\sim10\%$ relative to the average exposure within the plotted region.

\begin{figure*}
    \centering
    \includegraphics[width=\textwidth]{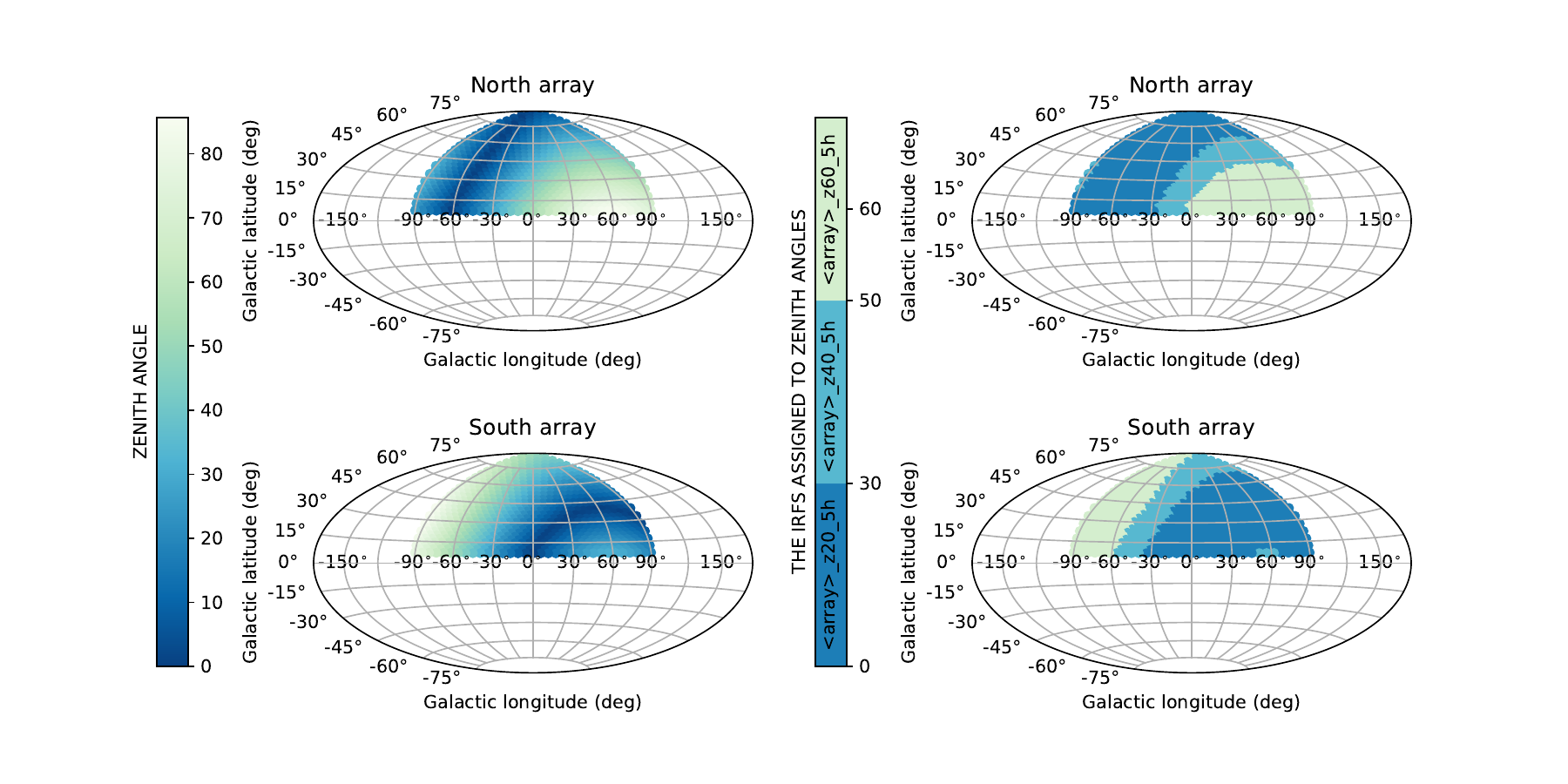}
    \caption{Strategy for assigning Instrument Response Functions (IRFs) across the extragalactic survey region as a function of zenith angle. In the left panel, the color scale shows, for each sky position, the zenith angle as measured in the observatory reference frame during observation (the values of the zenith angles and their color coding are reported on the vertical bar on the left). Since the IRFs are derived in three zenith angle bins (indicated by the color bar on the side of the right panel), the right panel shows which IRF is assigned to each region of the sky. The three IRFs shown here refer to 5 hrs of observation and to zenith angles below 30 deg (dark blue), between 30 and 50 deg (light blue) and above 60 deg (green).
}
    \label{fig:zenith_irf}
\end{figure*}

The IRFs we use are optimized for observations under different zenith angles. {For each pointing direction, we assign one of the three available IRF optimizations based on the minimum zenith angle at which that direction can be observed.} To estimate the minimal zenith angle, we use the Earth coordinates of both array positions, i.e., $(l_N, b_N) = (17.8920, \, 28.7622)$ and $(l_S, b_S) = (79.4041,\, -24.6272)$ for the Northern and Southern array, respectively. The zenith angle is calculated as the angular distance between the array location and the pointing direction when they are closest. The resulting zenith angles for the Northern and Southern arrays are shown in the left panel of \cref{fig:zenith_irf}. The strategy for assigning the IRFs based on zenith angle is adapted from Ref.\ \cite{Remy:2022}. For zenith angles smaller than $30^\circ$, we use the IRF optimized for $20^\circ$ angles; for zenith angles between $30^\circ$ and $50^\circ$ we use the IRF optimized for $40^\circ$ angles, and for zenith angles larger than $50^\circ$ we use IRF optimized for $60^\circ$ angles, as shown in the right panel of \cref{fig:zenith_irf}. All IRFs used in this work are optimized for 5 hours of observation time, independently of the array or the zenith angle.

The observational strategy presented in Sec. 8 of Ref. \cite{CTAConsortium:2018tzg} divides the extragalactic survey area into two parts: roughly $\sim$15\% of the sky {to be} observed using the Southern array and $\sim$10\% with the Northern array. {However, the precise division of the survey observations between the two arrays has not yet been finalized. The planned strategy remains under discussion and is still subject to adjustments.} For simplicity, we simulate observations of each pointing direction within the survey boundary using both Southern and Northern arrays with the corresponding zenith angle optimized IRFs. We use 1.11 hours and 0.51 hours per pointing direction with the Northern and Southern arrays, respectively. {This strategy provides a reasonable approximation for simulating the survey when combined with zenith-angle–optimized IRFs, as regions observed at higher zenith angles will yield fewer detected events, and vice versa.}  As a result, the sky coverage of both arrays will be complementary, as {illustrated} in the right panel of \cref{fig:zenith_irf}.

The survey observations are simulated for energies between 30 GeV and 30 TeV, using 30 log-spaced energy bins to ensure proper sampling of the IRFs across the energy range. For the purpose of estimating the background events and exposure over the whole extragalactic survey, we simulate the entire survey ROI, using coarser spatial binning of $1.0^\circ \times 1.0^\circ$ per pixel. For the sensitivity analysis, we instead simulate templates covering an ROI of $6.0^\circ\times 6.0^\circ$, using a finer spatial binning of $0.02^\circ \times 0.02^\circ$ per pixel.

\subsection{Off-source data scenario}

In addition to performing extensive surveys of the $\gamma$-ray sky, CTAO will be operated as an open, proposal-driven observatory. The observation time awarded to the wider research community can be added to the dedicated survey observation times of CTAO, and increasing the overall exposure that will be achieved over the first 10 years of telescope operation. An educated guess based on the performance of current IACTs can be made for CTAO observations performed through the proposal-driven program. In {Ref.} \cite{Coronado_Blazquez_2021}, the authors calculate the sky coverage and resulting observation time of the CTAO proposal program, assuming similarity with the actual operations of the MAGIC telescope. They extrapolate the 6.5 years of data obtained with the MAGIC array to 10 years of observation time using the two CTAO arrays. We adopt a similar approach to estimate the most optimistic case scenario for the observations of the extragalactic sky with CTAO, assuming an effective observation time of $\sim50$ hours per point and the same sky coverage as in the EGAL survey, consistent with the findings in {Ref.} \cite{Coronado_Blazquez_2021}. In the sensitivity analysis described below, we use a corresponding re-scaling factor to estimate the point source sensitivity of CTAO {under this scenario}. Details on how we derive the effective observation time for this scenario can be found in {\cref{app:expo}}.

\section{The cross-correlation signal}
\label{sec:formalism}

The cross-correlation between a map of the UGRB, $\gamma(\hat{\bm n})$, and fluctuations in galaxy number counts, $g(\hat{\bm n})$, can be measured through the cross-correlation angular power spectrum (APS) \cite{camera2013novel,camera2015tomographic,fornengo2014particle}. In our analysis, the field $g$ will be the distribution of galaxies, which is a biased tracer of the underlying dark matter distribution. 
In the Limber approximation \citep{1953ApJ...117..134L,1992ApJ...388..272K,1998ApJ...498...26K}, the APS can be written as~{\cite{camera2013novel,camera2015tomographic,fornengo2014particle}}
\begin{equation}
 C_\ell^{\gamma g}=\int \frac{\de\chi}{\chi^2}\,\,W_{\gamma}(\chi)\, W_g(\chi)\,P_{\gamma g}\left(k=\ell/\chi,\chi\right)\;,
\label{eq:clfin}    
\end{equation}
 where $\ell$ denotes the multipole, $k$ the wavenumber, $\chi(z)$ is the radial comoving distance, related to redshift $z$ in a flat cosmology through the relation $\de\chi=c\,\de z/H(z)$ with $H(z)$ being the expansion rate of the Universe. The term $P_{\gamma g}(k,z)$ represents the 3D cross-correlation power spectrum (PS), embedding the information on the fluctuations of the two correlated fields. The function $W_{i=\{\gamma,g\}}$ denotes the window functions of the gamma-ray intensity field, $\gamma$, and the galaxy number counts fluctuations, $g$, {respectively, and describes the redshift distribution of each field} (see {\cref{app:model}} for full details).
 
As astrophysical $\gamma$-ray sources we consider only BL Lacs, subdivided into the two sub-categories of high-synchrotron-peaked (HSP) blazars and low and intermediate synchrotron-peaked (LISP) blazars \cite{DiMauro:2013zfa}.
We model the spectral energy distribution (SED) of these sources as power-laws with index $\Gamma$ and exponential cutoffs $E_{\rm cut}$, whose specific parameter values are reported in \cref{table:BLLACS}, together with the minimum and maximum luminosities of the population.
Since we focus on CTAO and thus on energies above approximately {20} GeV, 
other typical $\gamma$-ray emitters like misaligned AGN, star-forming galaxies, and flat-spectrum radio quasars can be neglected, since their contribution are dominant at lower energies \cite{Fornasa_2015}. Full details on the modeling of the astrophysical sources are provided in {\cref{app:model}}, together with the expressions for their window functions and 3D cross-correlation power spectra, obtained within the halo model formalism.

Let us just recall here that the window function of the $\gamma$-ray emission of astrophysical sources depends on the $\gamma$-ray luminosity function (GLF) of each specific source-type and on their SED, and is integrated over the luminosity of the sources up to a threshold luminosity, which corresponds to the minimal luminosity that a source must posses in order to be resolved by the detector (in this way, the window function which is then obtained pertains to the {\it unresolved} sources, those contributing to the UGRB). The full expression and the detailed modeling can be found in Appendix \ref{app:Wastro}.

\begin{figure}[t!]
    \centering
    \includegraphics[width=0.45\textwidth]{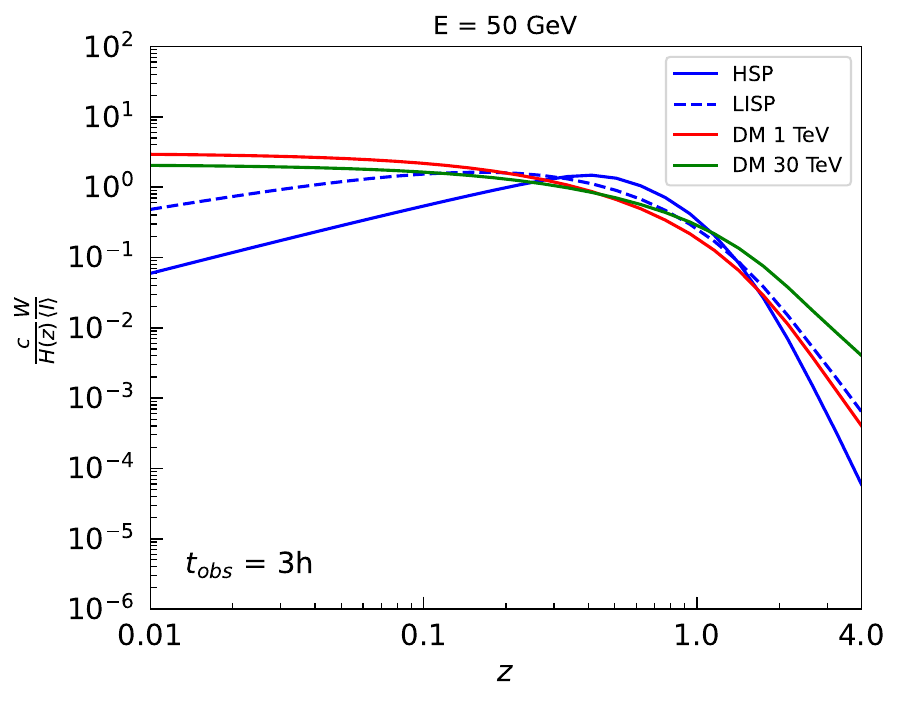}
    \includegraphics[width=0.45\textwidth]{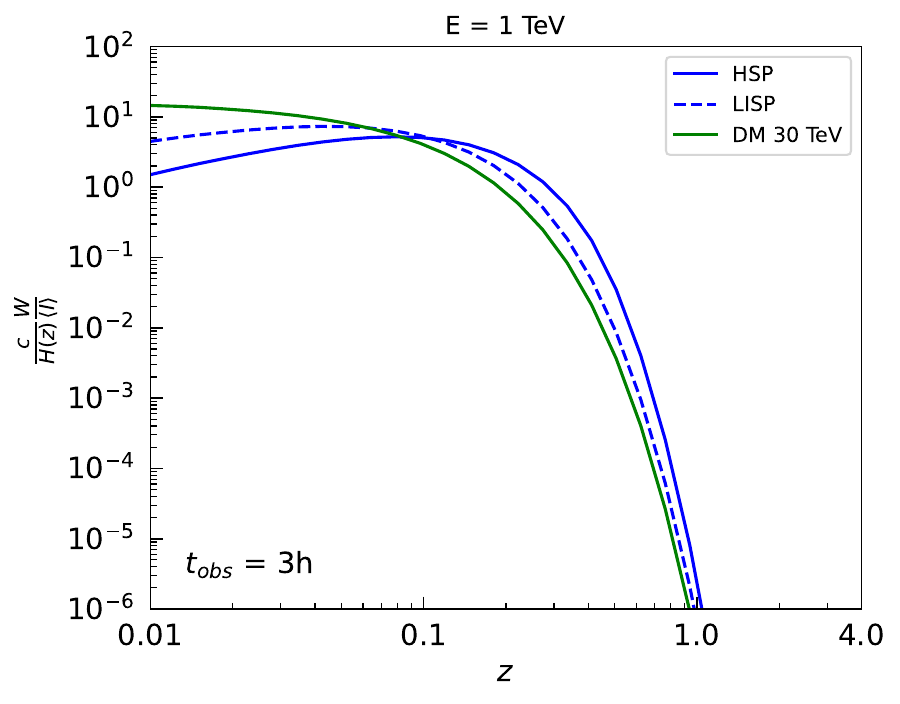}
    \caption{The $\gamma$-ray window functions, normalized to the mean $\gamma$-ray intensity and divided by the Hubble parameter, as a function of redshift. Blue solid and dashed lines show HSP and LISP astrophysical sources for 3 hours of observation, respectively. Red and green lines correspond to DM annihilation into $b\bar{b}$, $\langle \sigma v \rangle = 3 \times 10^{-26}$cm$^{-3}$s$^{-1}$ for DM masses of 1 TeV (red) and 30 TeV (green). Upper panel: $E_\gamma = 50$ GeV; lower panel: $E_\gamma = 1$ TeV (only 30 TeV DM case shown).}
    \label{fig:window_main}
\end{figure}

For the DM $\gamma$-ray emission, we consider both annihilating and decaying DM. In the case of annihilation, the relevant particle physics parameters are the DM mass $m_\chi$ and the thermally-averaged annihilation cross-section $\langle \sigma v\rangle$. For decaying DM, the key parameters are the mass and the lifetime $\tau$. These will be the parameters for which we provide the forecast sensitivity in the cross-correlation analysis. The {expressions for the} window functions and the 3D cross-correlation power spectra are reported in {\cref{app:model}}. 
Here we recall that the window functions are proportional to $\langle \sigma v\rangle$ for annihilating DM and inversely proportional to $\tau$ in the case of decaying DM, and proportional to $m_\chi^{-2}$ and $m_\chi^{-1}$ for annihilating and decaying DM, respectively. The window function of the gamma-ray emission from DM annihilation or decay also depends on the average dark matter content of the Universe and on the energy spectrum produced by dark matter annihilation or decay, which in turn depends on the specific annihilation/decay channel, for which we will consider $b \bar b$ as the representative case and provide results also for the channels $e^+e^-$, $\tau^+\tau^-$ and $W^+W^-$. In the case of annihilating dark matter, the window function also contains a so-called flux-multiplier factor, which accounts for the enhancement of the flux due to the clumping of DM and the presence of sub-structures within the DM halos. The full expressions and the detailed modeling of the DM window functions can be found in Appendix \ref{app:WDM}. A few examples of the redshift behavior of the $\gamma$-ray window functions for DM annihilation are shown in Fig. \ref{fig:window_main}, for two photon energies and two DM masses. For comparison, we also show the window functions of unresolved LISP and HSP astrophysical source classes, assuming an observation time of 3 hours. The observation time affects the window functions of astrophysical sources, since it determines the minimum luminosity of the sources that can be resolved by the instrument (and therefore do not contribute to the unresolved component consider in our analysis). The case for 50 hours of observation is shown in Fig. \ref{fig:zWindows} in {\cref{app:model}}.

For the present analysis, which primarily focuses on the sensitivity to a DM signal, we adopt as galaxy catalogs the Two Micron All-Sky Survey (2MASS) \cite{2MASS:2006qir,Bilicki:2013sza} and the 2MASS Redshift Survey (2MRS) \cite{Huchra:2011ii,Ando:2017wff} catalogs, since they cover the low redshift range $z \lesssim 0.1$ where most of the DM $\gamma$-ray emission originates.
Astrophysical sources, on the other hand, have a peak in emission at redshifts $z\sim 0.3$--$0.4$ at $E\sim {50}\,\mathrm{GeV}$, and $z\sim 0.1$--$0.2$ at $E\gtrsim 1\,\mathrm{TeV}$
as shown {in Fig. \ref{fig:window_main} and in Fig. \ref{fig:zWindows}} in {\cref{app:model}}.
The latter range is a consequence of strong absorption of $\gamma$-rays above $\sim 1 \mathrm{TeV}$ due to interactions with the extragalactic background light (EBL), which limits the $\gamma$-ray horizon to $z\sim 0.1$--$0.2$ independently of the type of source.

The present analysis is thus optimized for DM searches, while for astrophysical sources it will be beneficial to consider galaxy catalogs covering higher redshifts, up to $z\sim 0.4-0.5$. We leave this extension to future studies. The window functions of galaxy number counts are given by the redshift distribution of the number of objects in the catalog. Full details on the window functions of the catalogs adopted here can be found in {\cref{app:Wgal}}. The modeling of galaxy clustering within halos is based on the halo occupation distribution (HOD) formalism~\cite{Zheng:2004id,Berlind:2001xk,Cooray:2002dia}, which is also outlined in {\cref{app:model}}, together with the 3D cross-correlation power spectra. 

Let us note that in our analysis, the window function $W_g(z)$ of the galaxy catalogs is normalized to unity: $\int \de z\, W_g(z) =1$. For the $\gamma$-ray window function $W_\gamma(\chi)$, we define it such that the (average) intensity energy spectrum is obtained as: $I_{\gamma}(E) = \int \de\chi\ W_\gamma(E,\chi)$. 

The cross-correlation APS signal has a variance $\smash{(\Delta C^{\gamma g}_\ell)^2}$ given by~{\cite{camera2013novel,camera2015tomographic,fornengo2014particle}}
\begin{multline}
    (\Delta C^{\gamma g}_\ell)^2 \,(2\,\ell+1)\,f_{\rm sky} =  \\ (C_\ell^{\gamma g})^2 + \left[C_\ell^{\gamma \gamma} + \frac{C_N}{(B_\ell)^2}\right]\,\left(C_\ell^{gg} + C_{N_g} \right)\;, \label{eq:Delta}
\end{multline}
where $C_\ell^{\gamma\gamma}$ and $C_\ell^{gg}$ are the $\gamma$-ray and galaxy auto-correlation APS, $C_N$ is the photon noise, $C_{N_g}$ the galaxy catalog noise, and $B_\ell$ denotes the beam correction {accounting for the finite angular resolution of the $\gamma$-ray instrument. No beam correction is applied to the galaxy term, since the angular resolution of galaxy catalogs (typically of the order of arcseconds) is negligible for the purposes of this analysis. The parameter} $f_{\rm{sky}}$ denotes the fraction of the sky covered by the analysis. All the relevant terms in \cref{eq:Delta} are outlined in {\cref{app:model}}.

For completeness, in the following analysis, we will also use the $\gamma$-ray auto-correlation APS, which is a measure of the intrinsic anisotropy of the $\gamma$-ray emission. It reads~\cite{Ando:2013ff,camera2013novel,camera2015tomographic,fornengo2014particle}
\begin{equation}
 C_\ell^{\gamma \gamma}=\int \frac{\de\chi}{\chi^2}\,W_{\gamma}^2(\chi)  
 \,P_{\gamma \gamma}\left(k=\ell/\chi,\chi\right)\;,
\label{eq:clfin}    
\end{equation}
where $P_{\gamma \gamma}$ is the 3D power spectrum of the $\gamma$-ray emission. The corresponding variance on $C_\ell^{\gamma \gamma}$ is~{\cite{camera2013novel,camera2015tomographic,fornengo2014particle}}
\be
(\Delta C_\ell^{\gamma \gamma})^2 = \frac{2}{(2\,\ell+1)\,f_{\rm sky}} \left[C_\ell^{\gamma \gamma} + \frac{C_N}{(B_\ell)^2} \right]^2\;.\label{eq:variance_gammagamma}
\ee

\begin{table}[t!]
\centering
\begin{tabular}{|l |l l l l|} 
\hline
  & $\Gamma$ & $E_{\text{cut}}$ [GeV] & $L_{\rm min}$ [erg/s] & $L_{\rm max}$ [erg/s]  \\ [0.5ex]
 \hline
 HSP & 1.86 & 910 $^{+1100}_{-450}$ & $10^{38}$ & $3\cdot 10^{45}$  \\ [0.5ex]
 LISP & 2.08 & 37 $^{+85}_{-20}$ & $10^{38}$ & $3\cdot 10^{46}$ \\ [1ex]
 \hline 
\end{tabular}
\caption{Parameters and uncertainties for the power-law with exponential cut-off SED models of HSP and LISP populations, including the minimum and maximum luminosity, adopted from Table 1 of Ref. \cite{DiMauro:2013zfa}.}
\label{table:BLLACS}
\end{table}

\section{Sensitivity Forecast}\label{sec:sens}

\subsection{Astrophysical sources}

As a first step, we determine whether the anisotropy and the cross-correlation signals from unresolved astrophysical sources are detectable. As a statistical figure of merit, we adopt the signal-to-noise ratio (SNR), defined, for the cross-correlation, as
\begin{equation}
    {\rm SNR} = \sqrt{\sum_{\ell, i} \left(\frac{C_{\ell, i}^{\gamma g}}{\Delta C_{\ell, i}^{\gamma g}}\right)^2}\;,
    \label{eq:SNR2}
\end{equation}
where the sum runs over all the multipoles $\ell$ considered in the analysis and all energy bins $i$. For the auto-correlation, the formula is analogous but using $C_{\ell, i}^{{\gamma \gamma}}$ and $\Delta C_{\ell,i}^{\gamma \gamma}$.

To be more specific, the energy bins adopted in our analysis are 16, logarithmically spaced in the 30 GeV – 30 TeV range.
The sum over multipoles extends from $\ell = 10$ up to $\ell = 1000$. The lower value $\ell = 10$ is adopted to avoid the largest angular scales (i.e., lowest multipoles),
where the $\gamma$-ray Galactic foreground is the dominant emission, and, if not properly modeled and removed, can create some artifact in the reconstructed signal, especially in the auto-correlation. This issue is, however, mostly of concern for a real data analysis like the one in \cite{Fermi-LAT:2018udj}, where we refer the reader for further details. In our forecast, starting the sum from $\ell=10$ or $\ell=1$ has a negligible impact on the final results.
At the same time, we do not consider very large multipoles, since the (energy-dependent) angular resolution of the $\gamma$-ray detector prevents us from accessing information on angular scales which are much smaller than its point-spread-function (PSF). The CTAO PSF has been determined through a Monte Carlo simulation of the detector and is shown in Fig. \ref{fig:PSF_Beam} in Appendix \ref{app:model}, together with its harmonic space transform $B_\ell$, which enters in the expression for the variance $\Delta C_{\ell, i}^{\gamma g}$ of Eq. (\ref{eq:Delta}). From Fig. \ref{fig:PSF_Beam} we can see that for multipoles above about 1000 the beam function $B_\ell$ sharply decreases, thus inflating the variance and making those large multipoles irrelevant for the SNR.

The results for 3 hours and 50 hours of observation time are shown in \cref{table:SNR}, for both LISP and HSP source classes, and for both the auto-correlation and cross-correlation APS with the 2MASS galaxy catalog. We find that a purely astrophysical signal from unresolved sources can reach at most (for 50 hours of observation) only a marginal evidence of ${\rm SNR} \simeq 0.9$ in the autocorrelation, and about $0.16$ in the cross-correlation case.   
These values of SNR correspond to the HSP blazars, whose hard spectra are better suited to the CTAO energy range. The SNR is even smaller for LISPs, whose softer spectra result in a poorer overlap with the CTAO sensitivity. 

As an aside comment, let us note that this is not necessarily a negative result: it simply reflects the fact that the expected sensitivity of CTAO to point-source detection would be good enough to resolve a large fraction of blazars, thereby reducing the auto- and cross-correlation APS of the {\sl unresolved} population. In this case, it would be actually convenient and sufficient to perform auto- and cross-correlation APS analyses directly on the catalog of resolved sources in order to study their anisotropy properties. 

\begin{table}[t!]
\centering
\begin{tabular}{l l l} 
 \hline
 SNR for $3\,\mathrm{hrs}$ & Auto-correlation & Cross with 2MASS \\ [0.5ex]
 \hline
 LISP &  0.042 &   $8.2\cdot 10^{-3}$ \\ [0.5ex]
 HSP &  0.31 &  0.1 \\  [1.5ex]
 \hline \hline
  SNR for $50\,\mathrm{hrs}$ & Auto-correlation & Cross with 2MASS \\ [0.5ex]
 \hline
 LISP &  0.18 &  0.022 \\ [0.5ex]
 HSP &  0.9 &  0.16\\ [1ex]
 \hline
\end{tabular}
\caption{The predicted signal-to-noise ratios for 3-hours and 50-hours observations of LISP and HSP source classes, obtained for both $\gamma$-rays auto-correlation and cross-correlation with the 2MASS galaxy catalog. 
}
\label{table:SNR}
\end{table}

\begin{table}[t!]
\centering
\begin{tabular}{l l l} 
\hline
 SNR $50\,\mathrm{hrs}$ low res & Auto-correlation & Cross with 2MASS 
 \\ [0.5ex]
 \hline
 LISP &  0.7 &   0.033 \\ [0.5ex]
 HSP &  5.1 &  0.41 \\  [1ex]
 \hline
\end{tabular}
\caption{Same as \cref{table:SNR} but for $50\,\mathrm{hrs}$-observations, assuming a source detection threshold approximately three times worse than in \cref{table:SNR}. This illustrates the strong dependence of the auto-correlation signal on the point source detection threshold.}
\label{table:SNR-mix}
\end{table}

However, we wish to stress that this result, i.e., the fact that CTAO sensitivity would be enough to resolve a significant fraction of blazars, 
strongly depends on the assumed point-source sensitivity of the detector, which for CTAO {remains subject to considerable uncertainty}  (as discussed in \cref{sec:CTA}). This is because the {final configuration and specifications of CTAO have not yet been fully settled.}  For example, the point-source sensitivity {differs significantly between}  CTAO-North and CTAO-South \cite{CTAConsortium:2018tzg}, and at the moment the relative share of North and South array observations to the survey is still undecided. To assess the capabilities of CTAO for the study of anisotropies of point-sources even in the case that the point-source sensitivity is worse than the nominal case derived and adopted in this paper, we repeat the analysis for point-sources by assuming a degraded point-source detection capability. Table \ref{table:SNR-mix} shows the SNR obtained for 50 hrs of observations, assuming a source detection threshold approximately three times worse than in the nominal case of Table \ref{table:SNR}. This has been achieved by using the 50 hrs exposure with a point-source sensitivity corresponding to $3\,\mathrm{hrs}$ of observations, which is approximately a factor of 3 worse than the nominal $50\,\mathrm{hrs}$ sensitivity
(Note that, since we are in a strongly background-dominated regime, the uncertainty $\Delta C_{\ell}$ does not depend on the $C_{\ell}$ itself, but only on the background and the noise, as discussed in \cref{app:model}).
The result shows that the auto-correlation SNR for HSP indeed increases significantly, reaching ${\rm SNR} \simeq 5$.  The detectability of the cross-correlation APS signal also improves, but not so strongly as in the auto-correlation case. This is because the cross-correlation signal {is less sensitive to the point-source detection threshold. As detailed in the Appendices, the point-source sensitivity impacts the cross-correlation signal only through the window function, whereas it affects both the window function and the 3D power spectrum in the auto-correlation case.} 

Regarding LISPs, on the other hand, the SNR remains very low in both cases, reaching at most $0.7$ for the auto-correlation and $0.03$ for the cross-correlation. This is expected, because LISPs have a SED cut-off around $40\,\mathrm{GeV}$, resulting in minimal emission in the energy range where CTAO is most sensitive. 

\begin{figure*}[t!]
    \centering
    \includegraphics[width=0.45\textwidth]{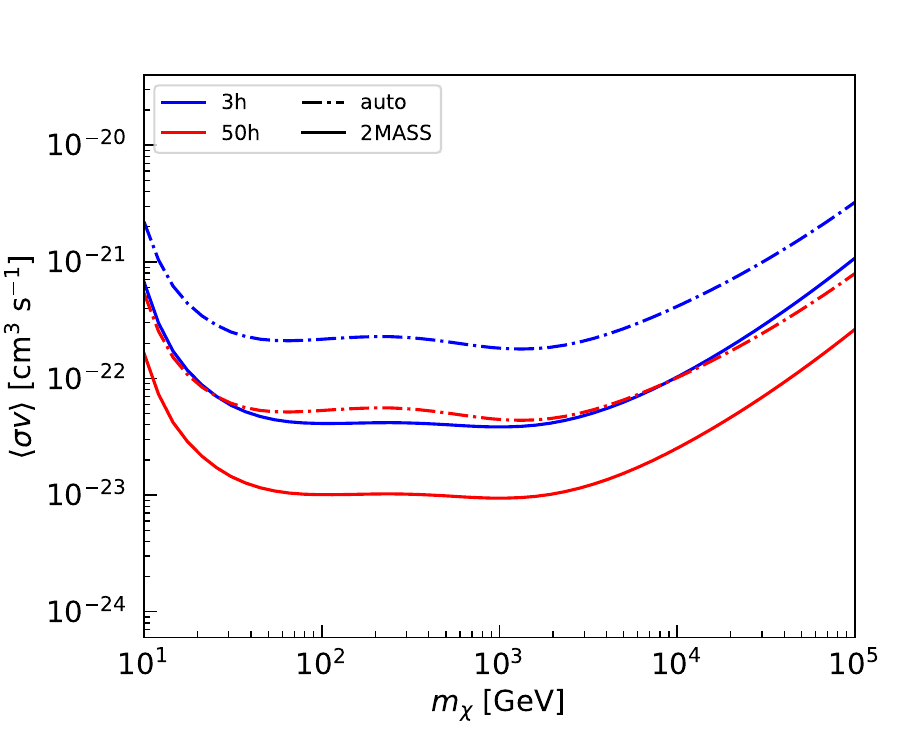}
    \includegraphics[width=0.45\textwidth]{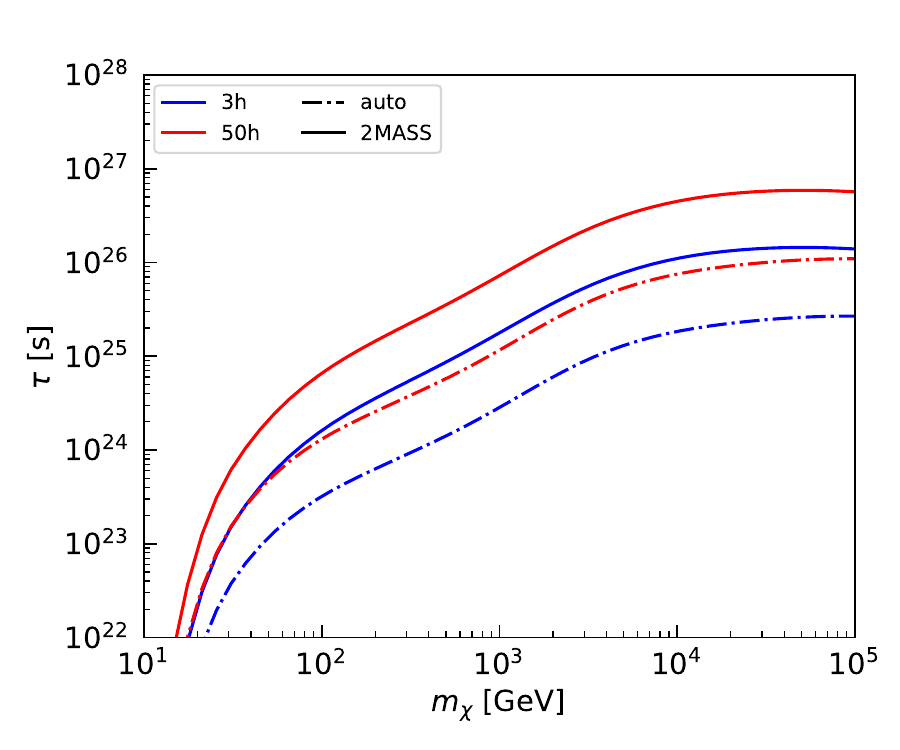}
    \caption{Left: Predicted upper bounds on the dark matter cross-section $\langle \sigma v \rangle$, for annihilation into $b\bar{b}$, based on 3-hour and 50-hour of observations. The curves represent the bounds from $\gamma$-ray auto-correlation (dot-dashed) and cross-correlation with the 2MASS galaxy catalog (solid). Right: Same as the left panel, but for the lower bound on the DM particle lifetime $\tau$.}
    \label{fig:bounds1}
\end{figure*}

\begin{figure*}[t!]
    \centering
    \includegraphics[width=0.45\textwidth]{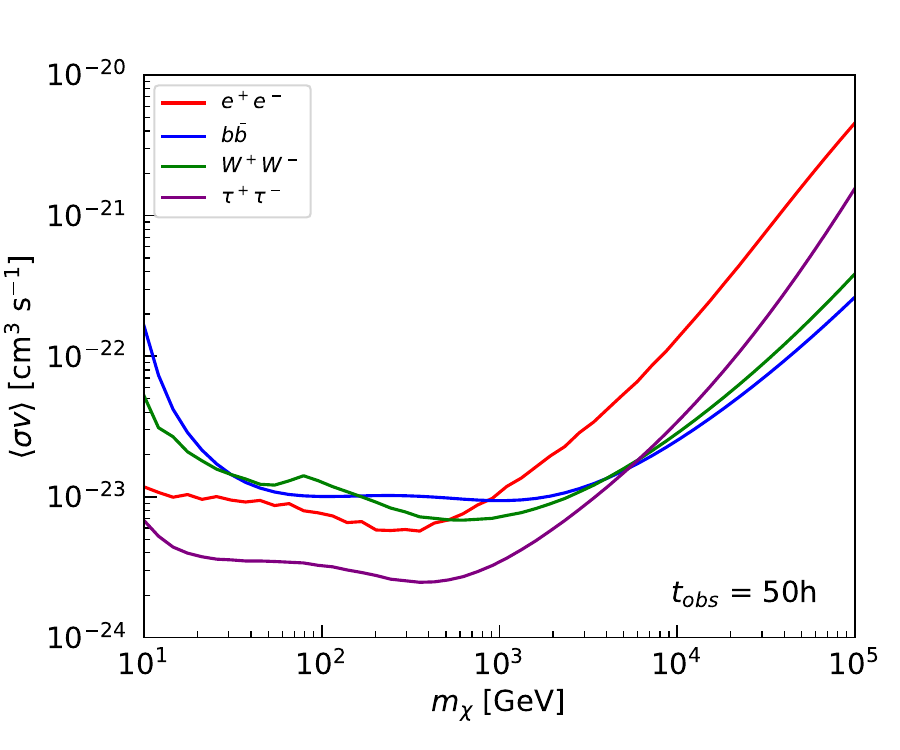}
    \includegraphics[width=0.45\textwidth]{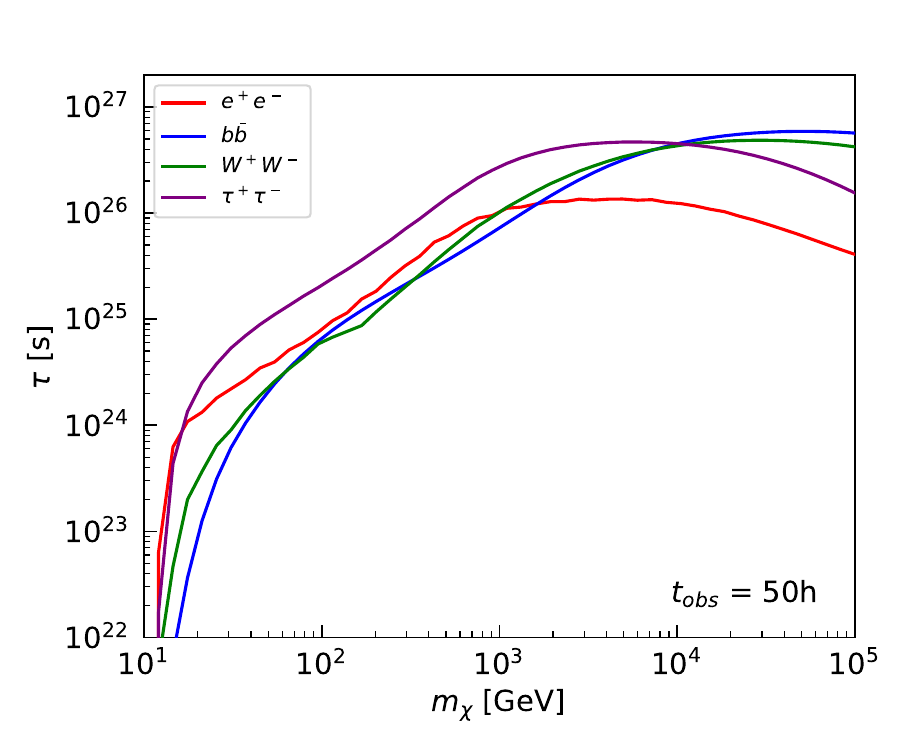}
    \caption{Left: Predicted upper bounds on the dark matter cross-section $\langle \sigma v \rangle$ from the cross-correlation with 2MASS galaxies, based on 50 hours of observation, for various dark matter annihilaton final states: $e^+e^-, b\bar{b}$, $W^+W^-$ and $\tau^+\tau^-$. Right: Same as the left panel, but for the lower bound on the DM particle lifetime $\tau$.}
    \label{fig:all_channels1}
\end{figure*}

\begin{figure*}[t!]
    \centering
    \includegraphics[width=0.45\textwidth]{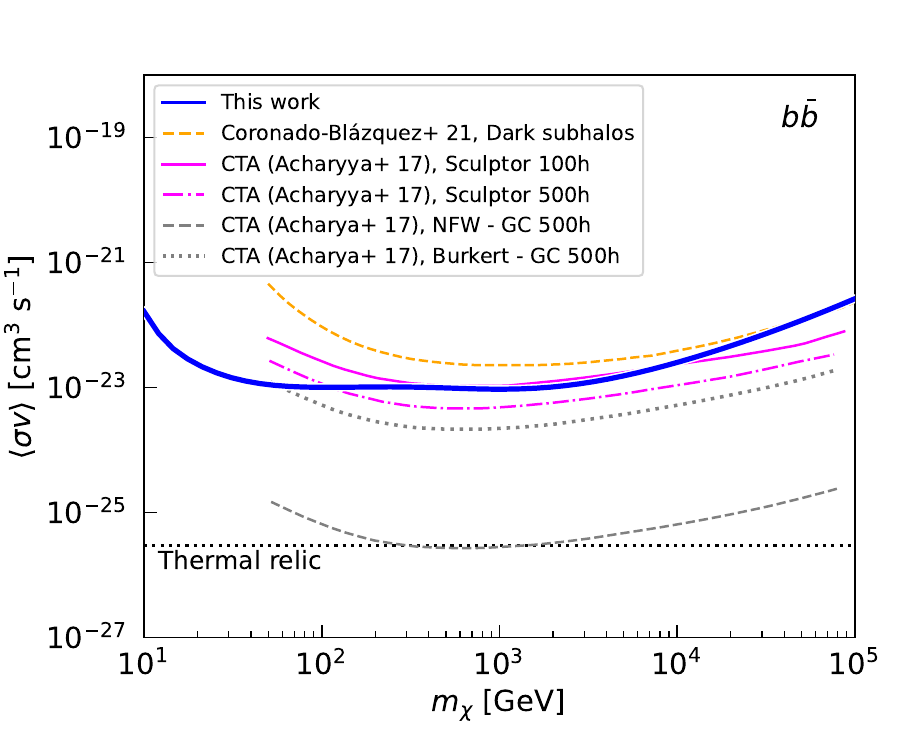}
    \includegraphics[width=0.45\textwidth]{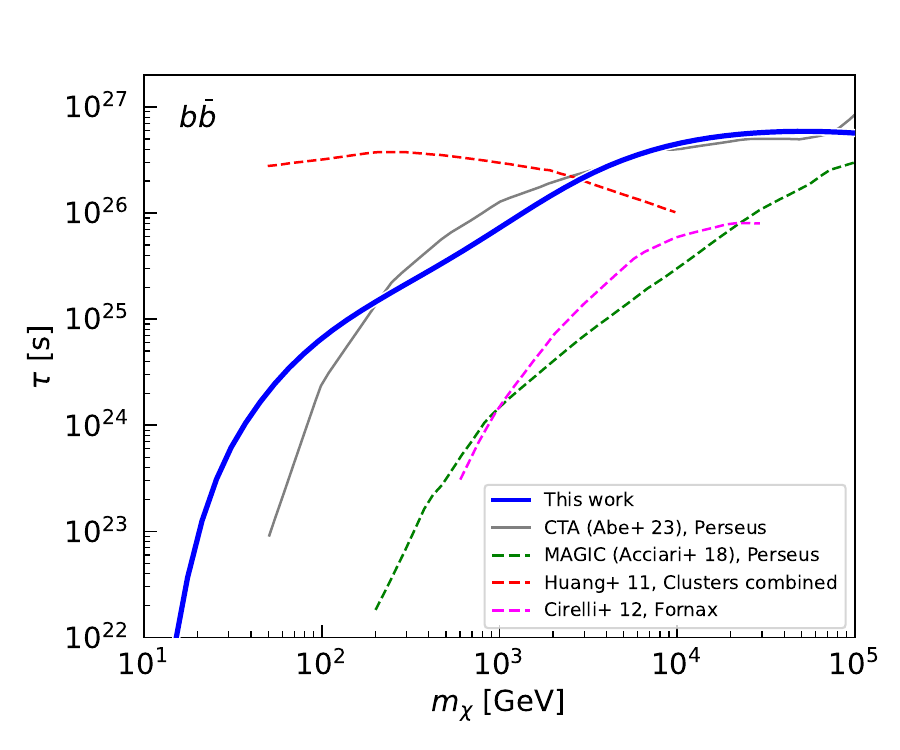}
    \caption{Left: Comparison between the forecast on the upper bounds on the annihilation cross-section $\langle\sigma v \rangle$ obtained in this work with other bounds from the literature, for the $b\bar{b}$ annihilation channels: bounds from non-observation of dark sub-halos (dashed yellow) \cite{Coronado_Blazquez_2021}, from Sculptor dwarf galaxy \cite{CTA_2017} with 100 hours (solid magenta) and with 500 hours (dot-dashed magenta) of observations with CTAO. Sensitivities from the Galactic halo have large uncertainties due to the assumptions on the dark matter profile toward the Galactic Center, therefore we show two limiting cases of an NFW profile (dashed gray) and of a cored profile (dotted gray) profiles for 500h with CTAO \cite{CTA_2017}. Right: Comparison between the forecast on the lower bounds on the DM particle lifetime $\tau$ obtained in this work with other bounds from the literature, for the $b\bar{b}$ decay channel: bound from Perseus with CTAO \cite{CTA2023} and MAGIC \cite{MAGIC_2018}, from galaxy clusters \cite{Huang_2012} and from Fornax \cite{Cirelli_2012}.}
    \label{fig:comparison1}
\end{figure*}

\subsection{Dark matter}
We now turn to assessing the CTAO sensitivity to an auto- or cross-correlation APS originating from DM annihilation or decay. To this end, we compare the null hypothesis, where only the contribution from astrophysical sources is present, with the alternative hypothesis scenario in which a DM $\gamma$-ray component is added on top of the astrophysical signal. From this comparison, we derive the minimum DM signal required to be statistically distinguishable from the purely astrophysical case. 

For cross-correlations, we adopt the following test statistics:
\begin{equation}
    \Delta \chi^2 = \sum_{\ell, i} \left(\frac{C_{\ell, i}^{\gamma_{\rm astro+DM}g}}{\Delta C_{\ell, i}^{\gamma_{\rm astro+DM}g}} \right)^2 - \sum_{\ell, i} \left(\frac{C_{\ell, i}^{\gamma_{\rm astro}g}}{\Delta C_{\ell, i}^{\gamma_{\rm astro}g}} \right)^2\;,
\end{equation}
where $\smash{C_{\ell, i}^{ \gamma_{\rm astro}g}}$ denotes the cross-correlation signal due to the presence of only astrophysical sources, while $\smash{C_{\ell, i}^{ \gamma_{\rm astro+DM}g}}$ refers to the case where a DM-induced $\gamma$-ray emission is present. The corresponding standard deviations are $\smash{\Delta C_{\ell, i}^{ \gamma_{\rm astro}g}}$ and $\smash{\Delta C_{\ell, i}^{ \gamma_{\rm astro+DM}g}}$, respectively. Since the DM signal depends on the DM mass $m_\chi$ and scales linearly with the annihilation rate $\langle \sigma v \rangle$ (or inversely with the lifetime $\tau$ in the case of decay), as outlined in {\cref{app:model}}, we derive bounds on the annihilation (or decay) rate by scanning over the DM mass $m_\chi$.
We then derive the $2\,\sigma$ bounds on the annihilation (or decay)
rate by requiring $\Delta \chi^2 = 4$. A similar test statistic and procedure is employed for the auto-correlation case.
We stress that the DM results are derived assuming the nominal case for the astrophysical sources scenario (i.e., HSP+LISP) and point-source sensitivity.
Nonetheless, the LISP component has no impact on the DM results due to low-energy SED cut-off, as explained in the previous section.
We have also verified that the point-source sensitivity uncertainty discussed in the previous section has a negligible impact on the DM results. To this aim, we explicitly derived DM sensitivities for the two cases of nominal and degraded point-source sensitivity, finding indistinguishable results.

\Cref{fig:bounds1} shows the sensitivity to the DM annihilation and decay in the $b\bar{b}$ annihilation channel, derived from the auto-correlation and cross-correlation APS with the  2MASS catalog, for $3\,\mathrm{hrs}$ and $5\,\mathrm{hrs}$ of CTAO observations. The $5\,\mathrm{hrs}$ case yields approximately a factor of 4 improvement in sensitivity compared to the $3\,\mathrm{hrs}$ case, while the cross-correlation outperforms the auto-correlation by approximately a factor of 5. This is expected since in the auto-correlation, the DM signal competes with strong blazar auto-correlation, while in the cross-correlation, the blazar contribution is much lower, yielding a more favorable signal-to-background ratio.
Sensitivities obtained from the cross-correlation with the 2MRS catalog are very similar and are shown {\cref{app:model}}.
Note that in this case, the uncertainty on the point-source detection threshold does not play a significant role, since the DM signal is always diffuse.

\Cref{fig:all_channels1} shows the sensitivities of the cross-correlation technique for different annihilation channels, while \cref{fig:comparison1} provides a comparison with existing results in the literature for alternative DM search strategies. In particular, we show bounds from observations of dwarf galaxies \cite{CTA_2017} and dark sub-halos \cite{Coronado_Blazquez_2021} for DM annihilation, and from galaxy clusters for decaying DM \cite{CTA2023,MAGIC_2018,Huang_2012,Cirelli_2012}. Our results show that the cross-correlation technique is competitive, yielding constraints comparable to those from other methodologies. For comparison, the figure also includes the sensitivity for annihilation in the Galactic halo, for 500 hr of observations with CTAO \cite{CTA_2017}: in this case, the values of $\langle \sigma v \rangle$ that can be probed can be lower, but the reachable bounds are strongly dependent on the assumptions regarding the DM profile near the Galactic Center.

In {\cref{app:modelll}} we show a comprehensive set of plots for different hours of observations and DM channels.

\section{Summary and conclusions}\label{sec:conclusions}

The CTAO will be the next generation of ground-based $\gamma$-ray Cherenkov detectors, offering a unique opportunity to study high-energy phenomena in the TeV range.  In addition to targeted observations, CTAO is expected to carry out large-area sky surveys, making it particularly well-suited for anisotropy-based techniques such as auto-correlation and cross-correlation analyses.
In this work, we have investigated the sensitivity of future CTAO data to constrain the properties of unresolved blazars and DM signals using these methods.

We found, in particular, that the auto-correlation APS method is better suited for probing the signal from unresolved blazars, with its sensitivity strongly dependent on the effective {point-}source detection threshold. On the other hand, the cross-correlation APS technique, when used in combination with galaxy catalogs, offers very promising sensitivity to potential DM annihilation or decay signal, reaching values around $10^{-23}\,\mathrm{cm^{-3}\,s^{-1}}$ for annihilation cross-section $\langle \sigma v \rangle$ and $10^{27}\,\mathrm{s}$ for decay.
These sensitivities are comparable to and complementary with those obtained from existing strategies, such as observations of dwarf spheroidal galaxies or galaxy clusters.

In the future, these results can be further improved with upcoming, deeper, and more densely sampled galaxy catalogs, such as those from the European Space Agency's \textit{Euclid} satellite mission \cite{2024arXiv240513491E}, as well as by incorporating cross-correlations with other large-scale-structure tracers, such as weak lensing.

\section*{Data Availability}
The data that support the findings of this article are openly available on Zenodo~\cite{pinetti_2025_17704724}.

\medskip

\begin{acknowledgments}
EP is grateful for the hospitality of Perimeter Institute, where part of this work was carried out. Research at Perimeter Institute is supported in part by the Government of Canada through the Department of Innovation, Science and Economic Development and by the Province of Ontario through the Ministry of Colleges and Universities. 
AA acknowledges support from Generalitat Valenciana through the `GenT program', ref.: CIDEGENT/2020/055 (PI: B. Zaldivar).
AC and NF acknowledge support from the Research grant TAsP (Theoretical Astroparticle Physics) funded by INFN. AC acknowledges support from: Departments of Excellence grant awarded by the Italian Ministry of University and Research (\textsc{mur}), and Research
grant ``Addressing systematic uncertainties in searches for
dark matter'', Grant No.\ 2022F2843L, CUP D53D2300258 0006 funded by \textsc{mur}. SC acknowledges support from \textsc{mur}, PRIN 2022 `EXSKALIBUR – Euclid-Cross-SKA: Likelihood Inference Building for Universe's Research', Grant No.\ 20222BBYB9, CUP D53D2300252 0006, and from the European Union -- Next Generation EU. NF is supported by the European Union – Next Generation EU and by \textsc{mur} via the PRIN 2022 project n. 20228WHTYC. GZ acknowledges Slovenian Research Agency grants P1-0031, I0-0033, J1-60014 and the Young Researcher program, Slovenia. 
The authors gratefully acknowledge the computer resources at Artemisa and the technical support provided by the Instituto de Fisica Corpuscular, IFIC(CSIC-UV). Artemisa is co-funded by the European Union through the 2014-2020 ERDF Operative Programme of Comunitat Valenciana, project IDIFEDER/2018/048. This work was conducted in the context of the CTAO
‘Dark Matter and Exotic Physics’ Working Group.

\end{acknowledgments}

\bibliographystyle{apsrev}
\bibliography{biblio_CTAxcorr}

\appendix

\onecolumngrid

\section{Off-source data scenario}
\label{app:expo}
\begin{figure*}
    \centering
    \includegraphics[width=0.32\textwidth]{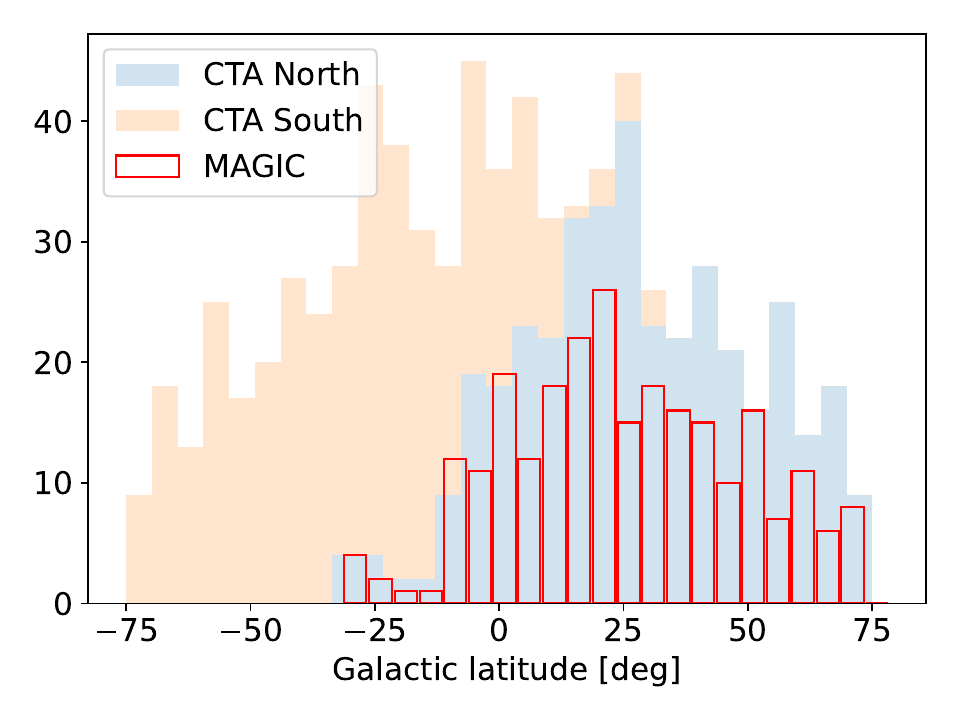}
    \includegraphics[width=0.32\textwidth]{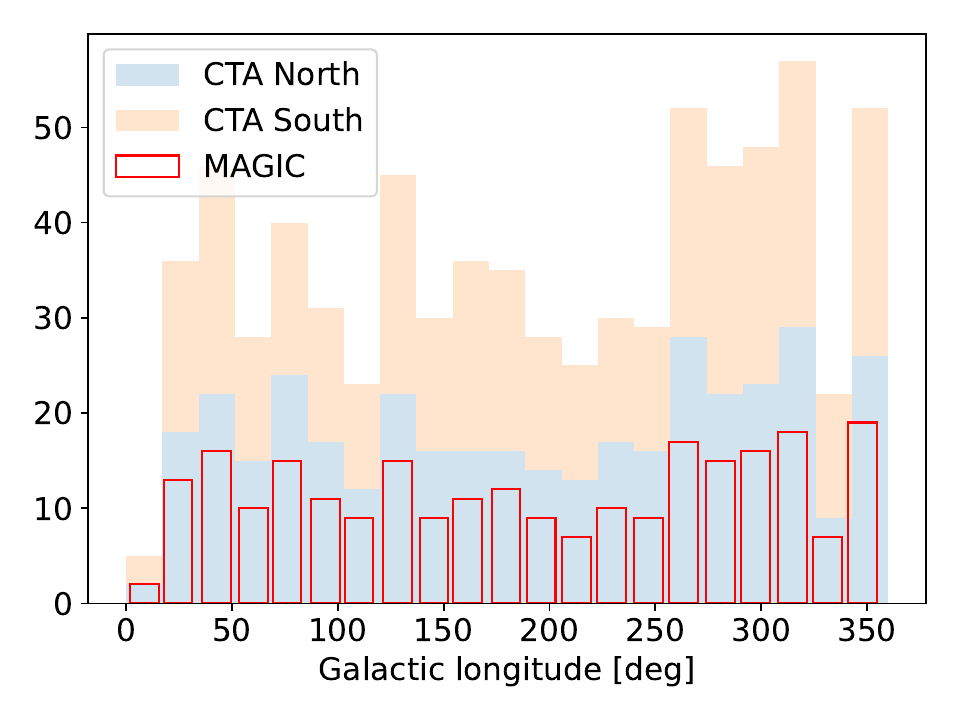}
    \includegraphics[width=0.32\textwidth]{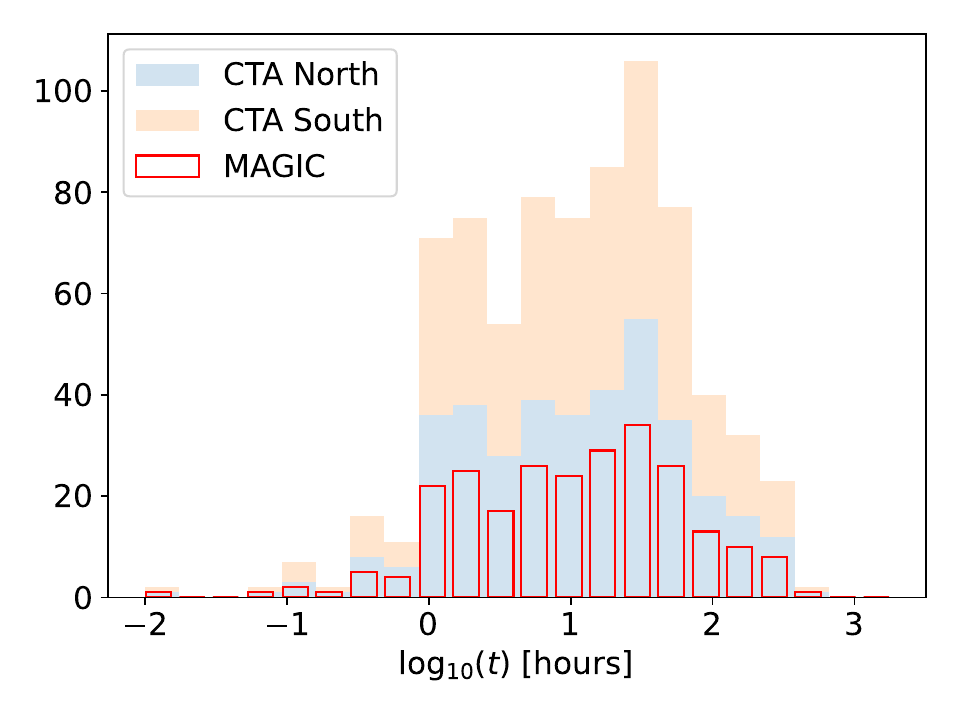}   
    \caption{Distributions of Galactic latitude (left), longitude (center), and observation time (right) for one realization of the predicted CTAO pointings over 10 years of observations. The stacked histograms show contributions from the Northern (blue) and Southern (orange) arrays. For comparison, the red outlined histograms indicate the corresponding distributions from 6.5 years of MAGIC observations.}
    \label{fig:expo_dist}
\end{figure*}

\begin{figure}[t!]
    \centering
    \includegraphics[width=0.45\textwidth]{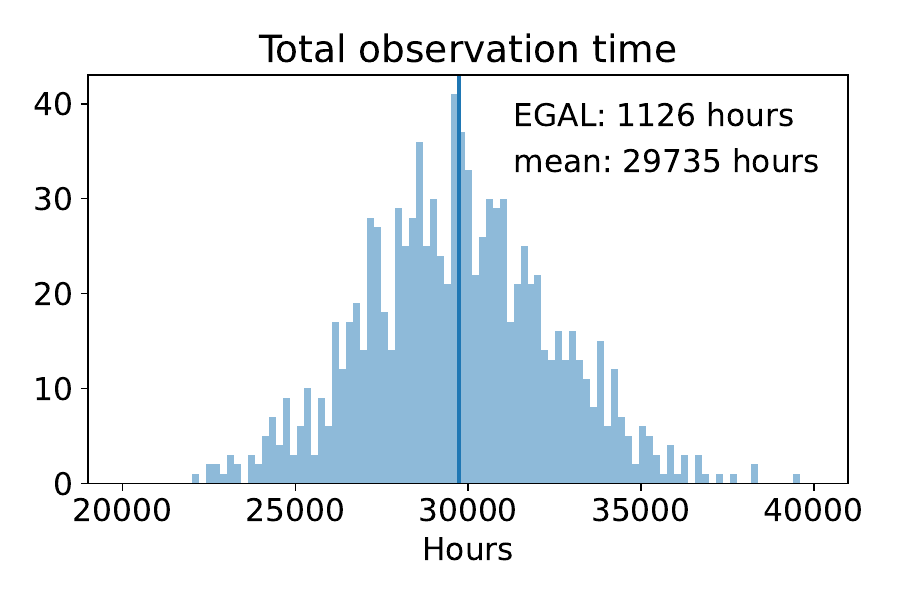}
    \caption{Distribution of the total observation time from 1000 realizations of predicted CTAO pointings across the whole sky using both arrays. The blue vertical line marks the mean total observation time. {The total time in the most optimistic scenario is about 25 times higher than that of the EGAL survey, effectively translating the nominal 3 hours per pointing into $\sim 50$ hours per sky location in this scenario.} }
    \label{fig:expo}
\end{figure}

As described in the main text, we follow the strategy outlined in Ref. \cite{Coronado_Blazquez_2021} to extrapolate MAGIC observations to CTAO. 
The distribution of MAGIC pointings, that we consider when extrapolating the observation time and pointing directions of future CTAO measurements across the sky, is shown in \cref{fig:expo_dist}, together with one realization of the predicted CTAO distributions of the same quantities, i.e. the Galactic latitude, longitude, and the observation time per pointing. The distribution of the cumulative pointing time over all pointings, obtained from 1000 different realizations of future CTAO observations, is shown in \cref{fig:expo}.

\section{Formalism and details of the modeling}
\label{app:model}

\subsection{Angular Power Spectra}

The angular power spectra for the $\gamma$-ray autocorrelation and for the cross-correlation between the unresolved $\gamma$-ray emission and a galaxy catalog $g$ can be expressed as \cite{camera2013novel,camera2015tomographic,fornengo2014particle}
\begin{eqnarray}
    \label{eq:auto}
     C_\ell^{\gamma \gamma} &=& \int \frac{\de\chi}{\chi^2}\,W_{\gamma}^2(\chi)  
 \,P_{\gamma \gamma}\left(k=\ell/\chi,\chi\right) \\ 
     C_\ell^{\gamma g} &=&\int \frac{\de\chi}{\chi^2}\,W_{\gamma}(\chi)\, W_g(\chi)\,P_{\gamma g}\left(k=\ell/\chi,\chi\right) \; , \label{eq:cross}
 \end{eqnarray}
 where $\ell$ denotes the multipole, $\chi$ is the comoving distance, $W_{\gamma}(\chi)$ and $W_g(\chi)$ are the window functions of the $\gamma$-ray emission and of the galaxy catalog, respectively, that is the functions describing how the observables are distributed in redshift.  $P_{\gamma \gamma}$ and $P_{\gamma g}$ are the 3D power spectra of the auto- and cross- correlation, respectively, which quantify the fluctuations of the corresponding observables. All these components are discussed in the following sub-sections.

\begin{table}[t!]
\centering
\begin{tabular}{|c|cc|}
\hline
Bin & $E_{\rm{min}}$ & $E_{\rm{max}}$ \\
\hline
 Low & 8 GeV & 126 GeV \\  
 Mid & 126 GeV & 1.3 Tev \\
 High & 1.3 TeV & 31.6 TeV \\
\hline
  \end{tabular}
\caption{Co-added energy bins used in the analysis for display purposes.}
  \label{Tab:energybin}
\end{table}

$C_\ell^{\gamma \gamma}$ and $C_\ell^{\gamma g}$ are both energy-dependent. In our approach, we compute the correlation functions within the specific energy bins for which the detector sensitivity is available. CTAO sensitivities are provided for 19 energy bins spanning the range (5 GeV, 501 TeV). Due to the sensitivity at low energies, we restrict our analysis to 16  bins, logarithmically spaced within the 30 GeV -- 30 TeV range. For visualization purposes, the energy-dependent quantities are further coadded into 3 broader bins, as summarized in \cref{Tab:energybin}. These low-, medium-, and high-energy bins correspond to the CTAO bin numbers 2-7, 8-12, and 13-18, respectively. However, when computing the DM constraints, we make sure of all 19 energy bins available for CTAO (bins 1 to 19), thereby covering the full energy range from 5 GeV to 501 TeV.

\begin{figure*}[t!]
    \centering
    \includegraphics[width=0.49\textwidth]{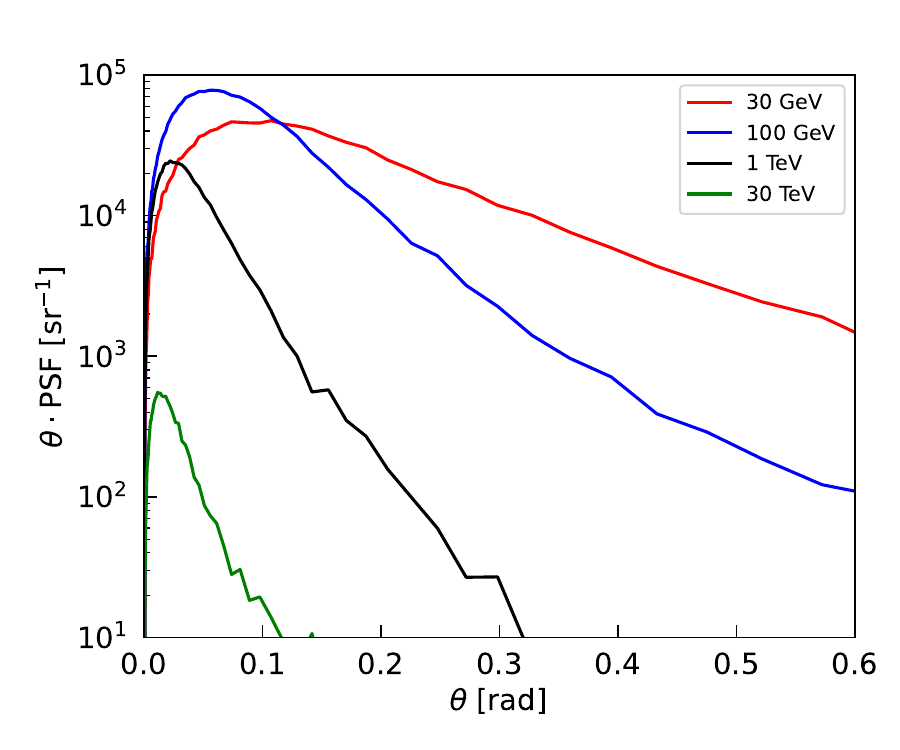}
    \includegraphics[width=0.49\textwidth]{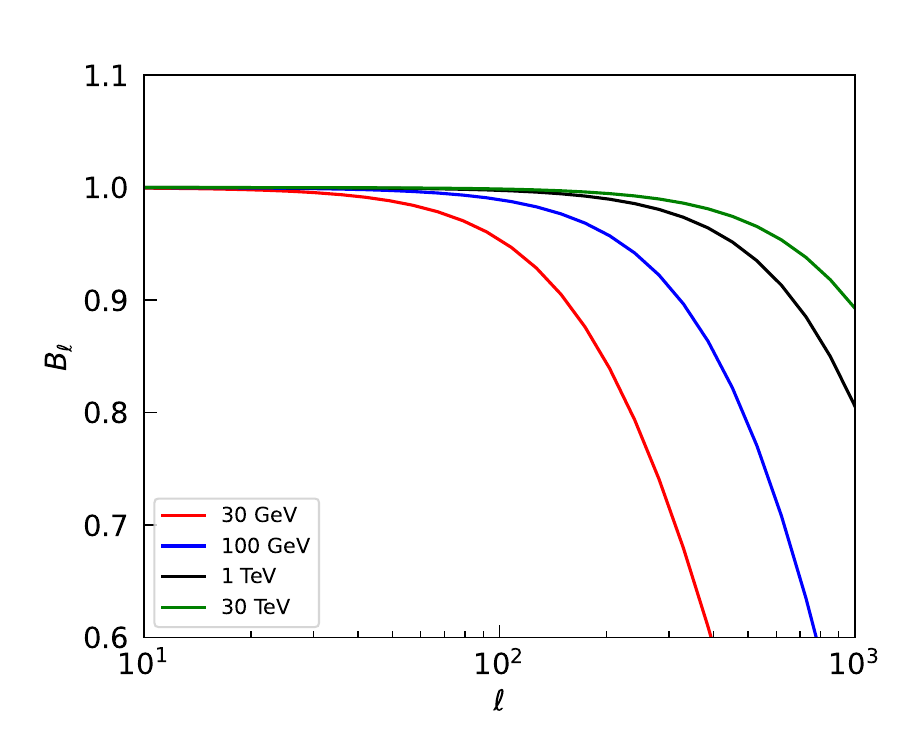}
    \caption{Left: Point spread functions of CTAO, as a function of angle for different photon energies. Right: The corresponding CTAO beam window functions as a function of multipole.}
    \label{fig:PSF_Beam}
\end{figure*}

The variance of the auto-correlation APS, in the Gaussian approximation, is given by \cref{eq:variance_gammagamma}.

The sky fraction observed by CTAO is $f_{\rm{sky}} \sim 0.25$ and it is considered energy-independent in our analysis. The beam window functions $B_\ell$ account for the smearing of the angular signal due to the finite angular resolution of the instrument and can be defined in terms of the point spread function (PSF) as 
\be
B_\ell(E) = \int \de\theta \, {\rm{PSF}} (\theta,E) \, \sin \theta \, P_\ell (\cos \theta)\;,
\ee
where $P_\ell (\cos \theta)$ are the Legendre Polynomials of order $\ell$. The point-spread function PSF($\theta, E $) has been obtained through a Monte Carlo simulation of the CTAO detector, and is energy-dependent.  \cref{fig:PSF_Beam} shows the PSF($\theta, E$) as a function of the angle $\theta$ (left panel) and the corresponding beam window function as a function of the multipole $\ell$ (right panel) for different energies.

The beam window function in a specific energy bin is then obtained by averaging over the energy-dependent intensity flux $\de I/\de E$, namely
\be
\langle B_\ell \rangle = \frac{1}{N}\,\int_{\rm bin} \de E \, B_\ell(E) \, \frac{\de I}{\de E}\;,
\ee
where $N = \int_{\rm bin} \de E \, {\de I/\de E}$ For definiteness, we use ${\de I/\de E} \propto E^{-2.3}$, which appropriately matches the observed behavior of the unresolved $\gamma$-ray intensity as measured by {\it Fermi}-LAT.

\bff
\centering
\includegraphics[width=0.49\textwidth]{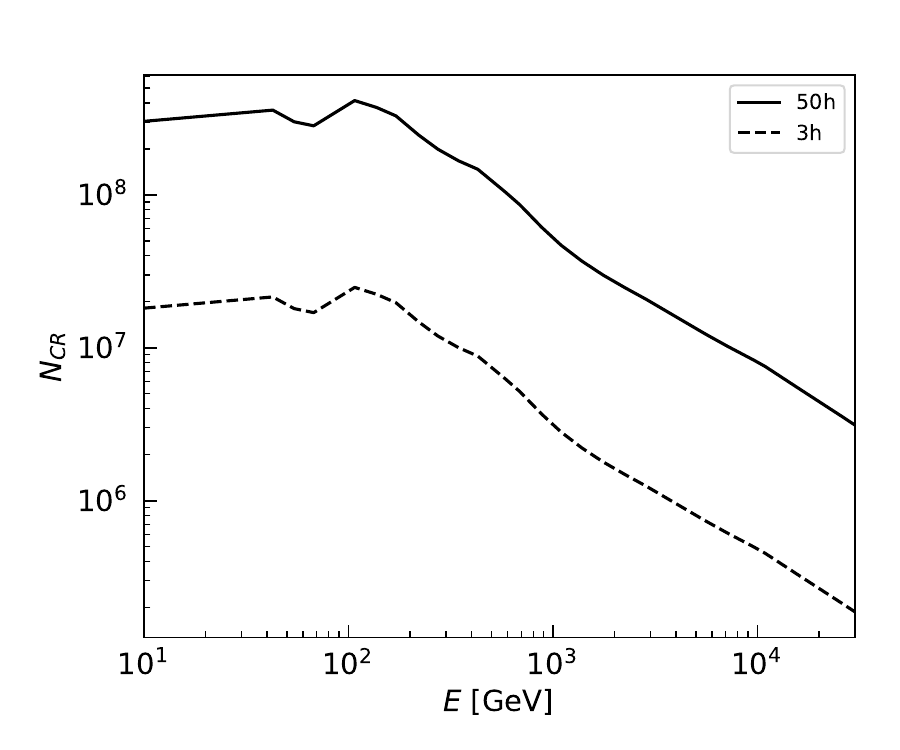}
\includegraphics[width=0.49\textwidth]{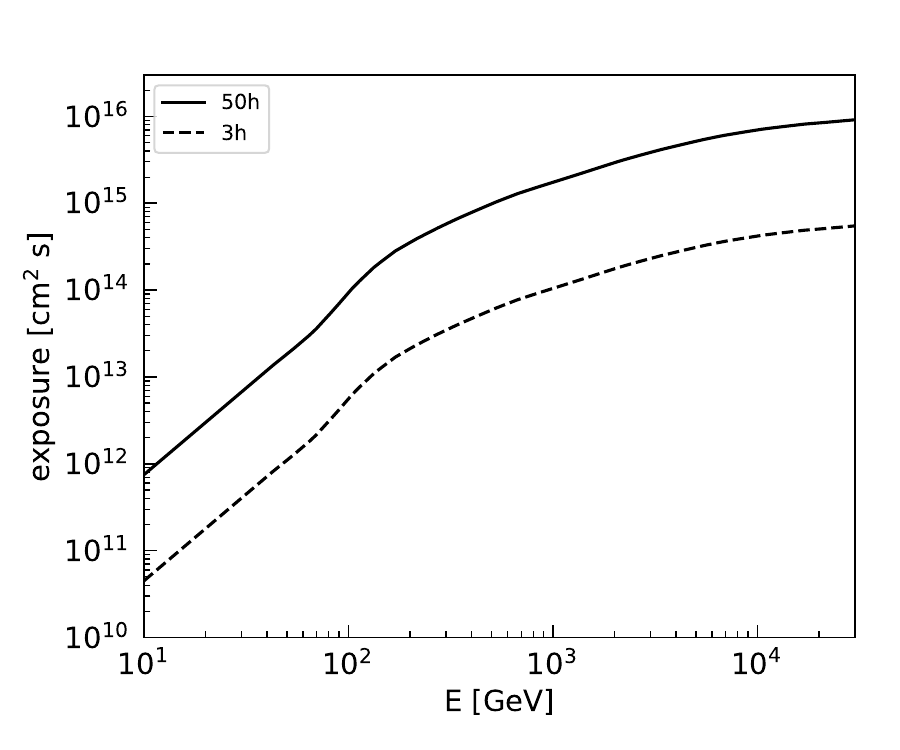}
\caption{Left: Cosmic-ray background $N_{\rm{CR}}$ as a function of energy. Right: CTAO exposure as a function of energy.}
\label{noise}
\eff

\begin{figure}[t!]
    \centering
    \includegraphics[width=0.45\textwidth]{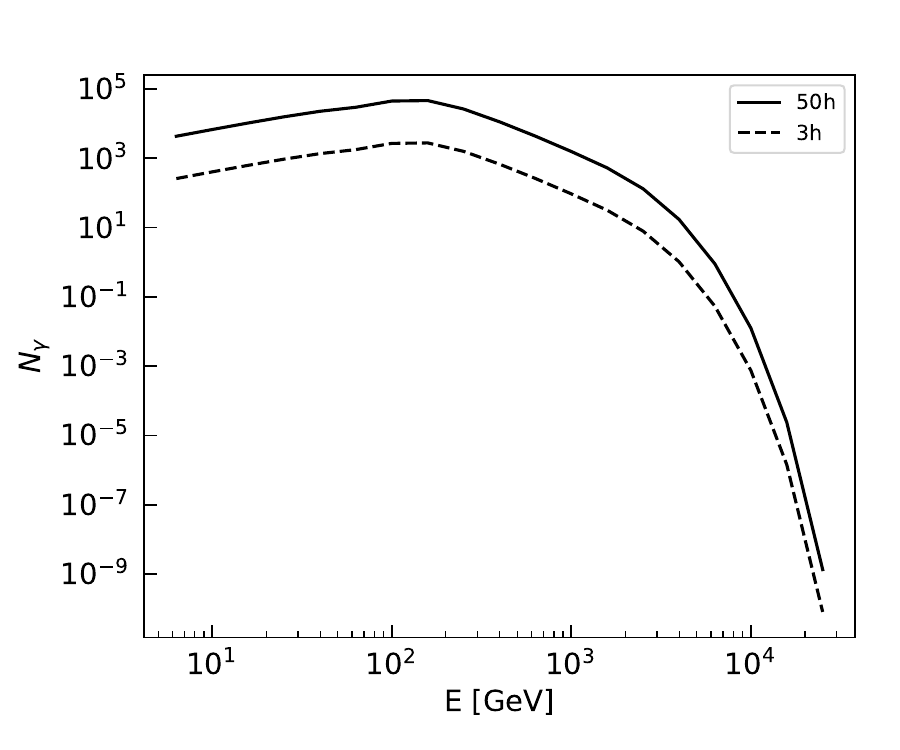}
    \includegraphics[width=0.45\textwidth]{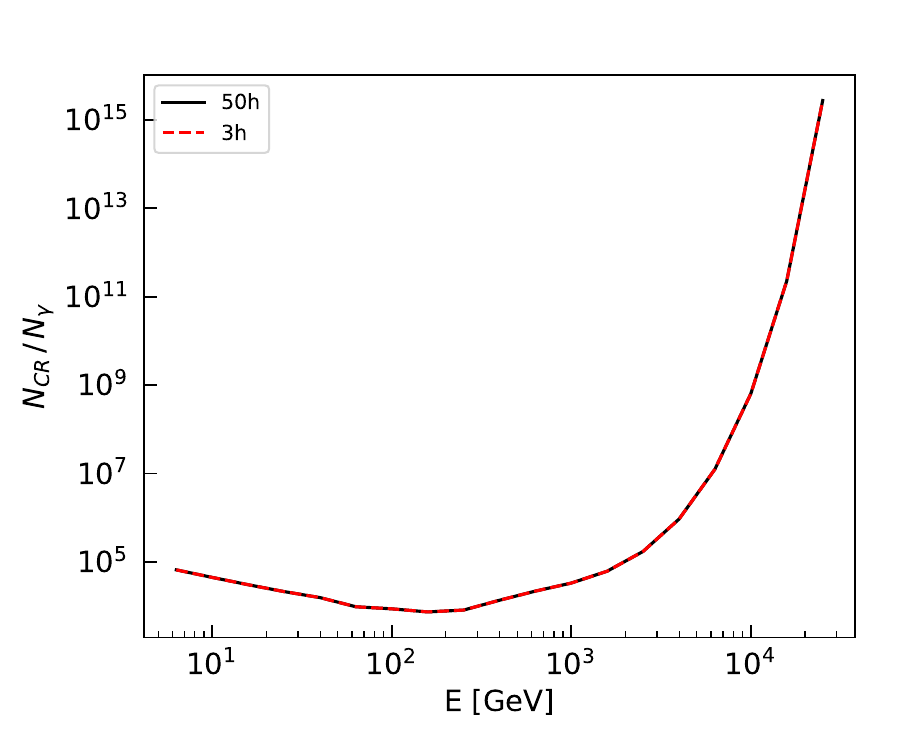}
    \caption{Left: Number of expected $\gamma$-ray events $N_{\gamma}$ in the CTAO survey as a function of energy. Right: Ratio {of cosmic-ray events to expected $\gamma$-ray events ($N_{\rm CR} / N_{\gamma}$)} after quality cuts in the CTAO survey. The ratio worsens above 1 TeV due to significant $\gamma$-ray suppression from EBL attenuation. $N_\gamma$ is determined for the HSP plus LISP source classes.}
    \label{fig:N_CR_over_N_gamma}
\end{figure}

The photon noise $C_N$ is also energy-dependent and is determined as
\be
C_N = \frac{4\,\pi\,f_{\rm sky}}{N_\gamma} \left(1 + \frac{N_{\rm{CR}}}{N_\gamma} \right)\,\langle I \rangle^2\;,
\ee
where $N_{\rm{CR}}$ represents the cosmic-ray irreducible background, $N_\gamma$ is the number of $\gamma$-ray events, and $\langle I \rangle$ denotes the mean $\gamma$-ray intensity in the specific energy bin, which is determined from the window functions as: 
$\langle I \rangle = \int \de\chi \,  W(\chi)$. $N_\gamma$ is therefore related to $\langle I \rangle$ via
\be
N_\gamma = 4\,\pi\,f_{\rm sky}\,\cal{E}\,\langle I \rangle\;,
\ee
where $\cal{E}$ represents the CTAO exposure in the given energy bin.
The exposure and the counts associated with the cosmic-ray background are derived via Monte Carlo simulations of the CTAO array. The simulation has been performed for 1.62h of observation time, and the results were then rescaled to obtain the exposure for $3\,\mathrm{hrs}$ and 50h. \cref{noise} displays the estimated cosmic-ray backgrounds $N_{\rm{CR}}$ (left panel) and exposures $\cal{E}$ (right panel) as a function of the energy.

 \cref{fig:N_CR_over_N_gamma}  shows the the number of $\gamma$-ray events $N_{\gamma}$ and the  ratio $N_{\rm{CR}}/N_{\gamma}$ as a function of energy. Since $N_{\rm{CR}}/N_{\gamma} \gg 1$  in all the energy bins considered in our analysis, the noise can be approximated as
\be
C_N \simeq 4\,\pi\,f_{\rm sky}\,\frac{ N_{\rm{CR}}}{N_\gamma^2}\,\langle I \rangle^2
\label{CN}
\ee
By substituting the expression of $N_\gamma$ in \cref{CN}, it becomes evident that $C_N$ can be expressed only in terms of detector properties and configuration,
\be
C_N \simeq \frac{ N_{\rm{CR}}}{4\,\pi\,f_{\rm sky} \, {\cal E}^2} 
\ee
The noise $C_N$ is shown explicitly in \cref{fig:Cl_vs_E}, together with a selection of auto- and cross-correlation energy spectra.

\begin{figure}[t!]
    \centering
    \includegraphics[width=0.9\textwidth]{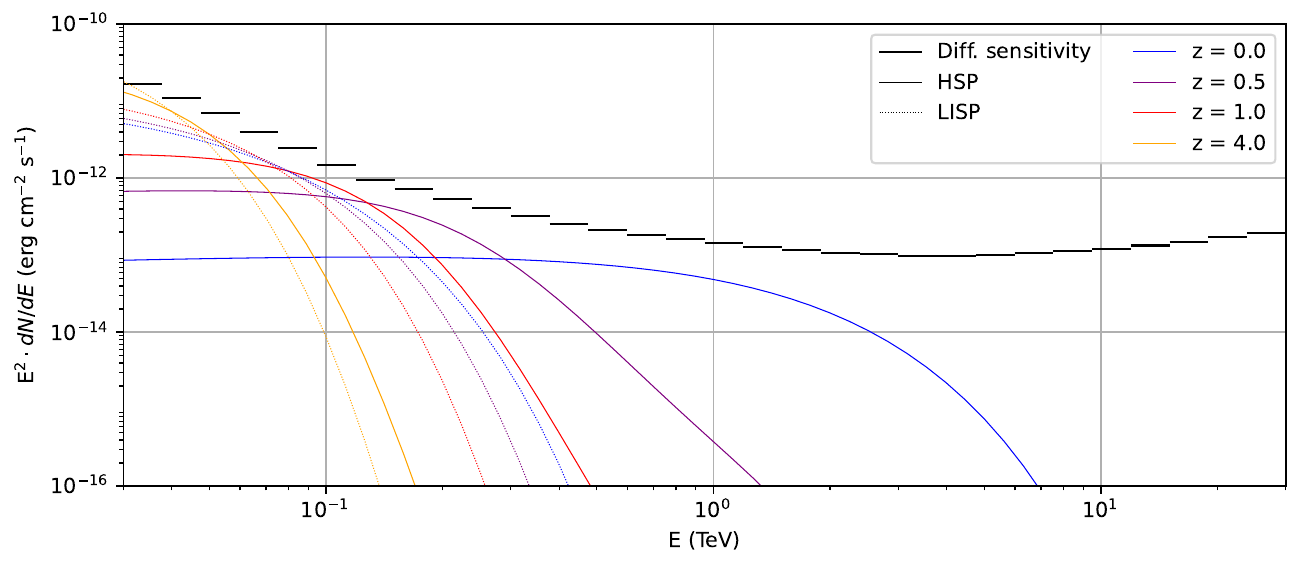}
    \caption{Differential flux sensitivity for $5\,\mathrm{hrs}$ of observation in the considered CTAO energy range (thick black horizontal dashes), compared with the energy spectra of HSP and LISP sources (solid and dotted lines, respectively) at four different redshifts. The spectra are normalized to the minimal flux detectable with the planned EGAL survey.}
    \label{fig:diffsens}
\end{figure}

In the case of cross-correlations, the variance is instead given by \cite{camera2013novel,camera2015tomographic,fornengo2014particle}
\begin{equation}
    (\Delta C^{\gamma g}_\ell)^2 = \frac{1}{\left(2\ell+1 \right) f_{\rm{sky}}} \left[(C_\ell^{\gamma g})^2 + \left(C_\ell^{\gamma \gamma} + \frac{C_N}{B_\ell^2}\right)\left(C_\ell^{gg} + C_{N_g} \right) \right].
    \label{eq:deltagamma g}
\end{equation}
where $C_\ell^{gg}$ is the galaxy auto-correlation, $C_N$ is the same as in the auto-correlation case, $C_{N_g}$ is the galaxy catalog noise. There is no beam correction in the galaxy term, as the angular resolution on the galaxy catalog is on the order of arcseconds,  making it negligible in this analysis. The galaxy noise is given by
\be
C_{N_g}= \frac{4\,\pi\,f_{{\rm sky},g}}{N_g}
\ee
where $f_{{\rm sky},g}$ represents the fraction of the sky covered by the catalog, and $N_g$ is the number of galaxies in the catalog. \cref{fig:Cl_vs_E} shows the quantity $\sqrt{C_N\,C_{N_g}}$ as a representative level for the cross-correlation noise, which approximately corresponds to the noise expected in a noise-dominated regime. 

\subsection{Window functions}
\label{app:window}

\subsubsection{Gamma-ray astrophysical sources}
\label{app:Wastro}

The window function for the $\gamma$-ray emission of unresolved astrophysical sources can be written as
\begin{equation}
    W(E, \chi) =  \frac{\de W}{\de E\,\de\chi} = \frac{d_L^2(z)}{\left(1+z \right)^2} \; \int_{L_{\rm min}}^{L_{\rm th}(z)} \;  \de L \; \Phi(L,z) \, \frac{d F}{d E}(E,L,z) \, \exp[-\tau(E,z)] 
    \label{eq:window_astro}
\end{equation}
where $d_L(z)$ is the luminosity distance, $\Phi(L,z)$ is the $\gamma$-ray luminosity function of the source population. In the present analysis, we use the luminosity functions of HSP and LISP as reported in \cite{DiMauro:2013zfa}.
Note that the above window function, which is used in Eq.\ref{eq:auto}-\ref{eq:cross} is differential in comoving distance, $\chi$. The analogous quantity differential, differential in $z$, would be $W(E,z)=\frac{\de W}{\de E\,\de z} = \frac{\de W}{\de E\,\de\chi} \frac{\de\chi}{\de z} = W(E,\chi) \frac{\de\chi}{\de z}$ where $\frac{\de\chi}{\de z} = \frac{c}{H(z)}$. Also note that the above window is differential in energy. In Eq. \ref{eq:auto}-\ref{eq:cross}, however, the corresponding quantity must be integrated over the energy bin considered. $L$ is the $\gamma$-ray luminosity of the source, conventionally integrated over the interval (0.1, 100) GeV, that is
\be
L = \int_{0.1\, {\rm GeV}}^{100\, {\rm GeV}} \de E_r \, \frac{\de L}{\de E_r}
\ee
where $E_r$ is the source rest-frame energy, related to the observed energy by $E_r = (1+z) E$, and $\de F/\de E$ is the spectral energy distribution (SED) of the sources, which we model as a power law with index $\Gamma$ and an exponential cutoff,
\begin{equation}
\frac{\de F}{\de E} = A(L,z) \; E^{-\Gamma} \, \exp(-E/E_{\rm cut})
\label{eq:SED}
\end{equation}
The values of $\Gamma$ and $E_{\rm cut}$ for the two classes of astrophysical sources adopted in our models are reported in \cref{table:BLLACS}, together with the minimum luminosity of the source classes $L_{\rm min}$. The parameter $A$, on the other hand, can be related to the luminosity $L$, by observing that (see Eq.\ 22 in \cite{Hogg:1999ad})
\be
L = \frac{4\,\pi\,d_L^2(z)}{(1+z)} \int_{0.1\, {\rm GeV}}^{100\, {\rm GeV}} \de E_r \, E_r \, \frac{\de F}{\de E_r} = 4\,\pi\,d_L^2(z) \int_{0.1\, {\rm GeV}/(1+z)}^{100\, {\rm GeV}/(1+z)} \de E \, E \, \frac{\de F}{\de E}
\label{eq:luminosity}
\ee
and therefore:
\be
A(L,z) = \frac{L}{4\,\pi\,d_L(z)^2} 
\left[ 
\int_{0.1\, {\rm GeV}/(1+z)}^{100\, {\rm GeV}/(1+z)} 
\de E \,  E^{(-\Gamma + 1)} \, \exp(-E/E_{\rm cut}) \right]^{-1}
\ee

The last term in \cref{eq:window_astro} models the $\gamma$-ray absorption on extragalactic background light (EBL). For the optical depth $\tau(E,z)$, we adopt the results from Ref. \cite{Finke:2009xi}. The maximal luminosity $L_{\rm th}(z)$ in \cref{eq:window_astro}, which refers to the threshold of detectability of resolved sources by the detector, depends on the CTAO sensitivity $F_{\rm sens}^{\rm CTA}(z)$ to point sources in the energy range of operation, which in our analysis is defined as (30 GeV, 30 TeV) 
\be
F_{\rm sens}^{\rm CTA}(z) = \int_{30\, {\rm GeV}}^{30\, {\rm TeV}} \de E \, \frac{\de F}{\de E} \, \exp[-\tau(E,z)] = 
A_{\rm th}(z)\; \int_{30\, {\rm GeV}}^{30\, {\rm TeV}} \, \de E \, E^{-\Gamma} \, \exp(-E/E_{\rm cut}) \exp[-\tau(E,z)]
\label{FsensCTA}
\ee
From $F_{\rm sens}^{\rm CTA}(z)$, it is therefore possible to determine the value of $A_{\rm th}(z)$, which allows us to determine the threshold luminosity by means of \cref{eq:luminosity} as
\be
L_{\rm th}(z) = 4\,\pi\,d_L^2(z) \; F_{\rm sens}^{\rm CTA}(z) \; 
\frac{
\int_{0.1\, {\rm GeV}/(1+z)}^{100\, {\rm GeV}/(1+z)} \de E \, E^{-\Gamma+1} \, \exp(-E/E_{\rm cut})}
{\int_{30\, {\rm GeV}}^{30\, {\rm TeV}} \, \de E \, E^{-\Gamma} \, \exp(-E/E_{\rm cut}) \exp[-\tau(E,z)]}
\label{eq:threshold}
\ee

\begin{figure}[t!]
    \centering
    \includegraphics[width=0.45\textwidth]{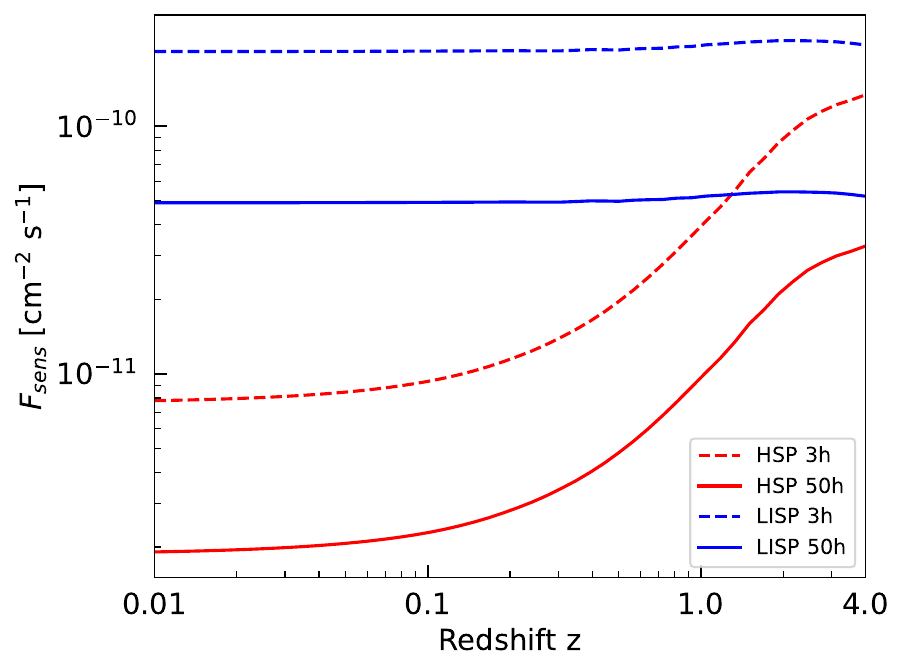}
    \caption{CTA sensitivity $F_{\rm sens}^{\rm CTA}(z)$ for $3\,\mathrm{hrs}$ and $5\,\mathrm{hrs}$ of observation as a function of redshift, for the HSP and LISP $\gamma$-ray source classes.}
    \label{fig:Fsens}
\end{figure}

The HSP and LISP point source sensitivity has been obtained through extensive Monte Carlo simulations \cite{CTAConsortium:2018tzg,VeronikaThesis} and is shown in \cref{fig:diffsens} for $3\,\mathrm{hrs}$ of observation across different energy bins. The same plot compares the differential sensitivity with the $\gamma$-ray spectra of LISP and HSP blazars normalized to the minimum detectable intensity. As shown, the spectra observed at Earth are redshift-dependent, as sources at different redshifts experience different levels of EBL absorption. \cref{fig:Fsens} instead shows the integrated HSP and LISP point source sensitivity in the energy range 30 GeV to 30 TeV as a function of $z$, where the normalization of the SED in \cref{eq:SED} has been fixed to its threshold value $A_{\rm th}(z)$, obtained from \cref{FsensCTA} by employing the Monte Carlo modeling of $F_{\rm sens}^{\rm CTA}(z)$.

\subsubsection{Gamma-ray emission from dark matter}
\label{app:WDM}

Window functions for dark matter are modeled as in Refs. \cite{camera2015tomographic,fornengo2014particle,Cuoco:2015rfa}, to which we refer for additional details. For decaying dark matter, we have
 \be
     W(E, \chi) = \frac{\de W}{\de E\,\de\chi} = \, \left (\frac{\Omega_{\rm DM} \,\rho_{\rm c} }{m_{\chi}} \right) \, \frac{\Gamma_{\rm d}}{4 \pi}  \frac{\de N}{\de E}[(1+z)/E] \, \exp[-\tau(E,z)]
     \label{eqn:Wann}
 \ee
where $\Omega_{\rm DM}$ is the dark matter density parameter, $\rho_{\rm c}$ is the critical density of the Universe,
$m_{\chi}$ the DM particle mass, $\Gamma_{\rm d}$ the DM decay rate, $dN/\de E$ the $\gamma$-ray energy spectrum per decay event, which we take from \cite{Arina:2023eic} for decay into $b\bar{b}$, $\tau^+ \tau^-$, $W^+ W^-$. The last term in the equation refers again to $\gamma$-ray absorption, as discussed for the astrophysical sources.

\begin{figure}[t!]
    \centering
    \includegraphics[width=0.45\textwidth]{figures/new_analysis/W_at_50GeV_3h.pdf}
    \includegraphics[width=0.45\textwidth]{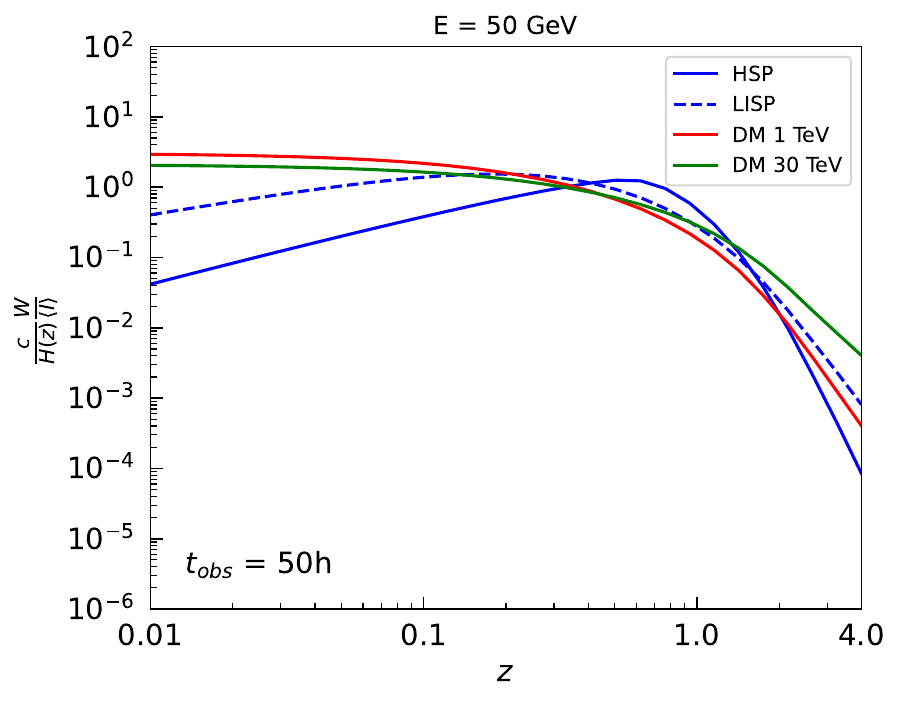}
    \includegraphics[width=0.45\textwidth]{figures/new_analysis/W_at_1TeV_3h.pdf}
    \includegraphics[width=0.45\textwidth]{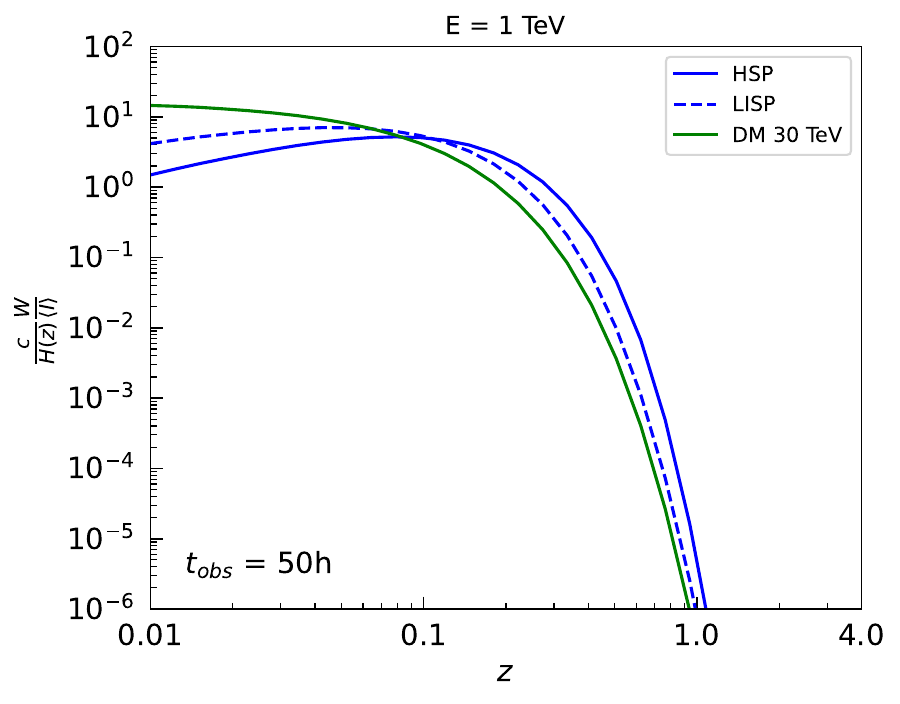}
    \caption{$\gamma$-ray window functions, normalized to the mean $\gamma$-ray intensity, as a function of redshift (and divided by the Hubble parameter). Left panels show results for 3 hours of observation, and right panels for 50 hours. The blue solid and dashed lines represent the HSP and LISP astrophysical sources, respectively. The red and green solid lines refer to DM annihilation into $b\bar{b}$, $\langle \sigma v \rangle = 3 \times 10^{-26}$cm$^{-3}$s$^{-1}$ and masses of 1 TeV (red) and 30 TeV (green). The upper row is calculated for a $\gamma$-ray energy of 50 GeV, while the lower row is for 1 TeV (only the 30 TeV DM case is shown here). {The 3-hour observation time cases are the same as shown in Fig. \ref{fig:window_main}}.}
    \label{fig:zWindows}
\end{figure}

\begin{figure}[t!]
    \centering
    \includegraphics[width=0.45\textwidth]{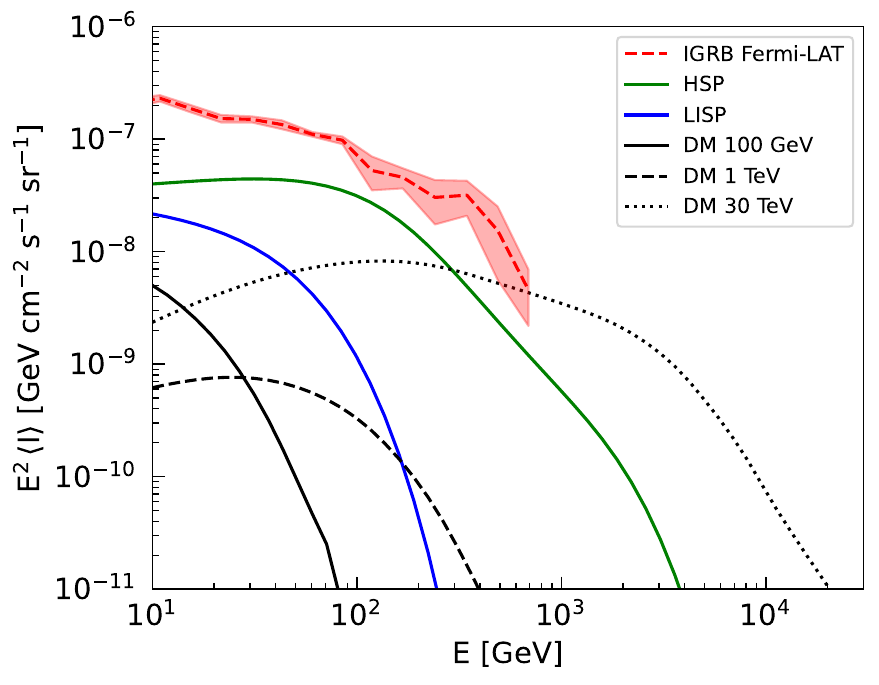}
    \caption{$\gamma$-ray intensity in our benchmark model for HSP (solid green) and LISP (solid blue), along with three cases of dark matter annihilation for particle masses of 100 GeV (solid black), 1 TeV (dashed black). and 30 TeV (dotted black). The 100 GeV and 1 TeV lines correspond to an annihilation cross-section $\langle \sigma v \rangle = 3 \times 10^{-26}$cm$^{-3}$s$^{-1}$, while for 30 TeV we adopted $\langle \sigma v \rangle = 3 \times 10^{-23}$cm$^{-3}$s$^{-1}$. The figure also shows the measured IGRB (red dashed line and red band for its uncertainty) \cite{Ackermann_2015}.}
    \label{fig:E2_mean_intensity}
\end{figure}

For annihilating DM, the window function reads
 \be
     W(E, \chi) = \frac{\de W}{\de E\de\chi} = \frac{1}{4 \pi} \, \frac{\langle \sigma v \rangle}{2} \Delta^2(z) \, \left (\frac{\Omega_{\rm DM} \,\rho_{\rm c} }{m_{\chi}} \right)^2 (1+z)^3  \frac{d N}{d E}[(1+z)E]  \exp[-\tau(E,z)]\;,
     \label{eqn:Wann}
 \ee
where  $\langle \sigma v \rangle$ is the velocity-averaged annihilation cross-section. Unlike DM decay, the annihilation window function also contains the flux multiplier factor $\Delta^2(z)$, which accounts for the enhancement of the flux due to the clumping of DM and the presence of sub-structures within the DM halos. Details on the way we model the flux multiplier can be found in Ref. \cite{Pinetti:2021jjs,Pinetti:2019ztr,Arcari:2022zul}.
Window functions for both DM and astrophysical sources are plotted in Fig.\ref{fig:zWindows} as a function of redshift for two different energies, and for $3\,\mathrm{hrs}$ and $5\,\mathrm{hrs}$ of observation time. It can be seen that the DM window always peaks at low $z$, while the astrophysical signal peaks at $z\sim 0.4-0.5$ at 50 GeV and  $z\sim 0.1-0.2$ at 1 TeV, with the horizon shrinking due to the effect of EBL $\gamma$-ray absorption.
\cref{fig:E2_mean_intensity}, instead, displays the intensity $\langle I \rangle$ as a function of the energy for the two classes of astrophysical sources used in the analysis and for a few representative cases of DM annihilation.

\subsubsection{Galaxy Catalogs}
\label{app:Wgal}

The galaxy-catalog window functions are given by the redshift distribution of the number of objects in the catalog, namely
\begin{equation}
      W_g(z) = \frac{\de N_g}{\de z}\;.
\end{equation}
 The window function is normalized such that $\int \de z \, W_g(z) = 1$. Note again that $W_g(\chi)$ entering the cross-correlation APS is related to the redshift-dependent window function as $W_g(\chi) = W_g(z) \frac{\de z}{\de\chi} = W_g(z) H(z)/c$, where $H(z)$ is the Hubble parameter and $c$ the speed of light.
 In this way we also have  $\int \de\chi \, W_g(\chi) = 1$.
We use the $dN_g/\de z$ as given in \cite{Huchra:2011ii,Ando:2017wff} for 2MRS and \cite{2MASS:2006qir,Bilicki:2013sza} for 2MASS.

\subsection{Galaxies Halo Occupation Distribution}
To model the clustering of galaxies within the catalog, we use the Halo Occupation Distribution (HOD) formalism (see  \cite{Zheng:2004id,Berlind:2001xk,Cooray:2002dia}).

The number of galaxies residing within a DM halo of mass $M$ is parameterized as
\begin{eqnarray}
\langle N_{\rm cen}(M)\rangle &=& \frac{1}{2}\left[1+{\rm erf}\left(\frac{\log M-\log M_{\rm th}}{\sigma_{\rm logM}}\right)\right]\;, \label{eq:HOD1}\\
\langle N_{\rm sat}(M)\rangle &=& \left( \frac{M}{M_1} \right)^\alpha \exp{\left(-\frac{M_{\rm cut}}{M}\right)}\;,\label{eq:HOD2}
\end{eqnarray}
where, for a halo of mass $M$, $\langle N_{\rm cen}(M) \rangle$ is the average number of galaxies at the center of the halo and $\langle N_{\rm sat}(M)\rangle$ represents the average number of satellite galaxies.
The parameters $M_{\rm th}, \ \sigma_{\rm logM}, \ \alpha, \ M_1, \ M_{\rm cut}$ characterize the HOD and we chose them according to 
\cite{Ando:2017wff} for 2MRS and \cite{Cuoco:2015rfa} for 2MASS.

\subsection{3D Power Spectra}
The 3D Power Spectra $P(k)$  (PS) have been calculated by using the Halo Model formalism \cite{Cooray:2002dia}. We have derived the autocorrelation 3D PS of decaying DM $P_\delta(k)$, annihilating DM $P_{\delta^2}(k)$ and astrophysical sources $P_\gamma(k)$, and the 3D cross-correlation PS among the same three categories of tracers: $P_{g\delta}(k)$ for the cross-correlation between galaxies and decaying DM $\gamma$-ray emission,  $P_{g\delta^2}(k)$ for the cross-correlation between galaxies and annihilating DM $\gamma$-ray emission, for the cross-correlation between galaxies and astrophysical sources $\gamma$-ray emission $P_{g\gamma}(k)$. The explicit formulas for the power spectra are reported in Ref. \cite{Cuoco:2015rfa}, to which we refer the reader for details. 

\begin{figure}[t!]
    \centering
    \includegraphics[width=0.45\textwidth]{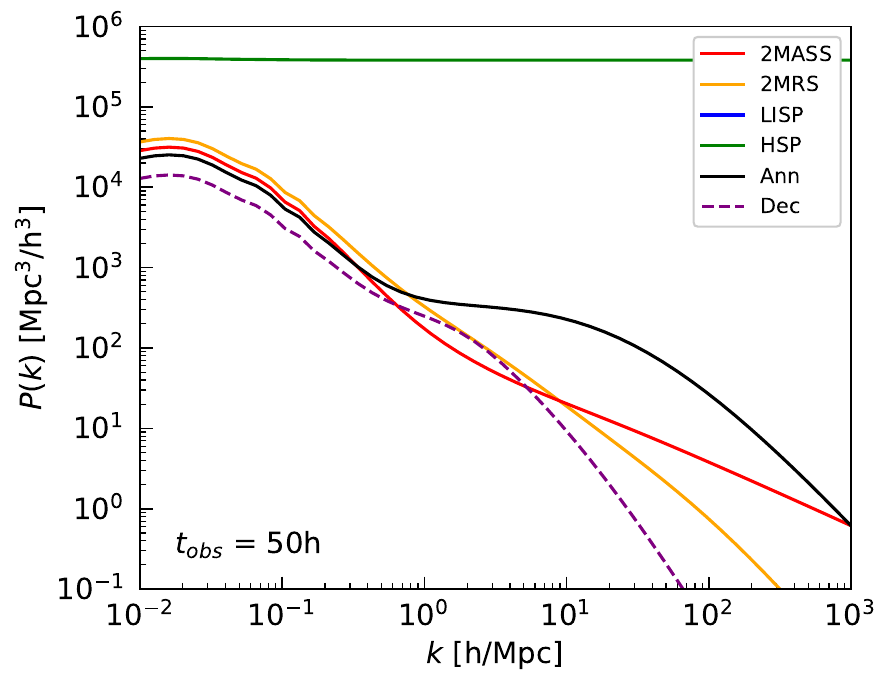}
    \includegraphics[width=0.45\textwidth]{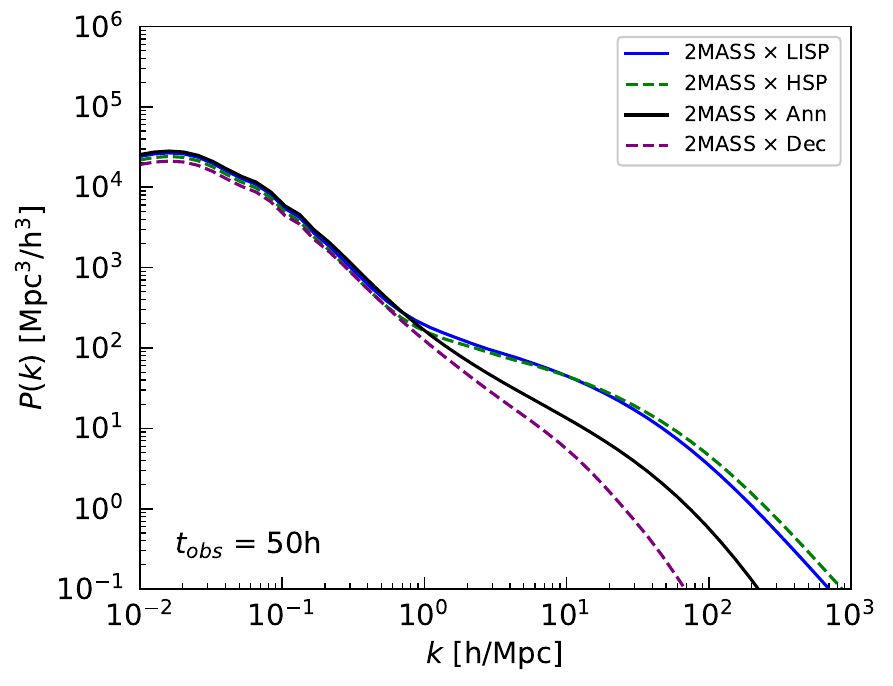}
    \caption{Left: Auto-correlation 3D power spectra for 2MASS and 2MRS galaxies, LISP and HSP astrophysical sources, and annihilating and decaying dark matter. Right: Cross-correlation 3D power spectra for the cross-correlation between 2MASS galaxies and the various $\gamma$-ray source types considered in our analysis: LISP, HSP, annihilating and decaying dark matter.}
    \label{fig:PS_tot}
\end{figure}

\begin{figure}[t!]
    \centering
    \includegraphics[width=0.45\textwidth]{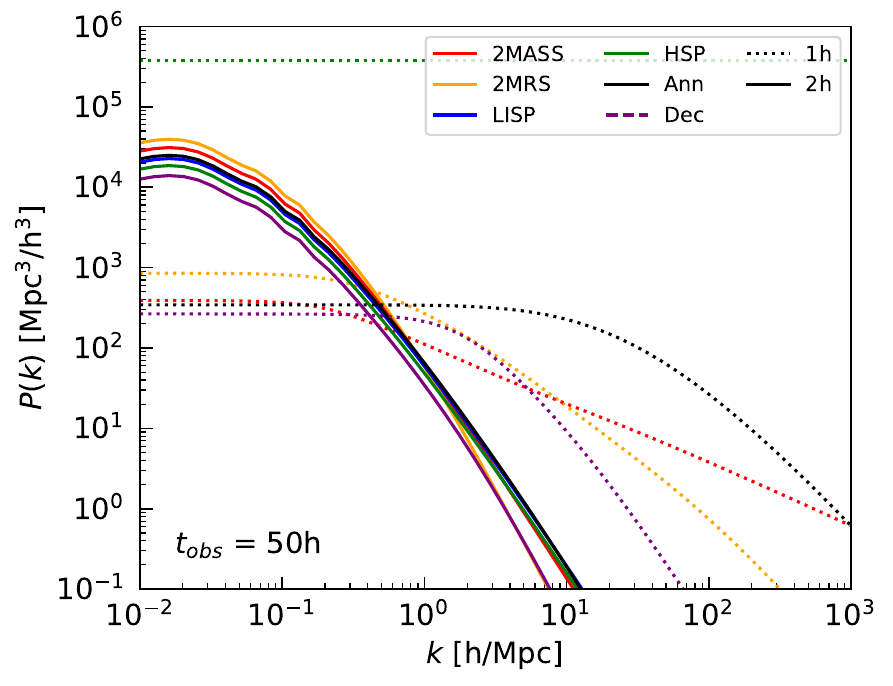}
    \includegraphics[width=0.45\textwidth]{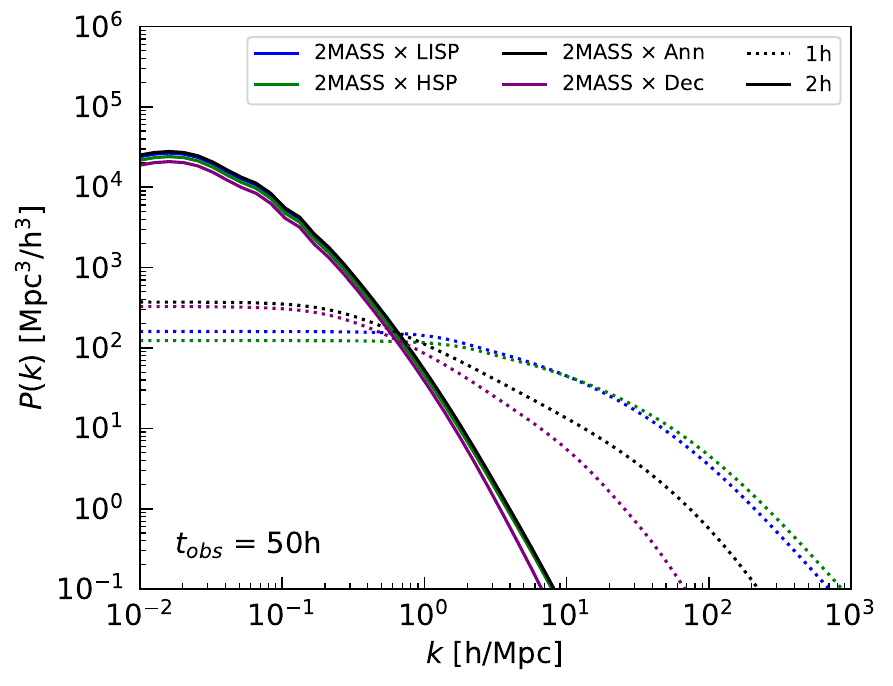}
    \caption{The same as Fig.\ref{fig:PS_tot} but with the 1-halo (dotted) and the 2-halo (solid) terms shown separately. }
    \label{fig:P1h_P2h}
\end{figure}

\begin{figure}[t!]
    \centering
    \includegraphics[width=0.45\textwidth]{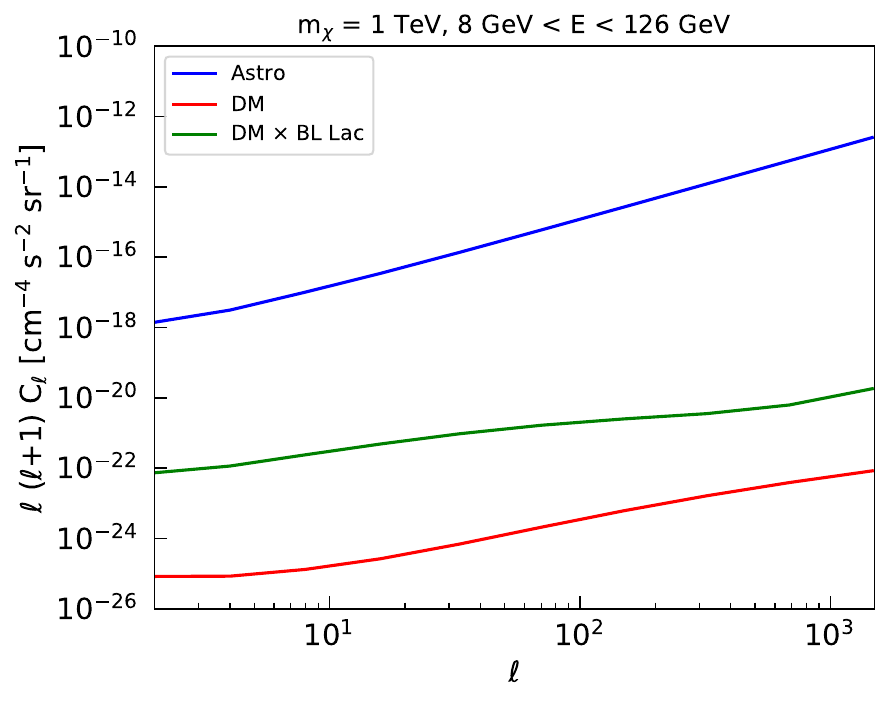}
    \includegraphics[width=0.45\textwidth]{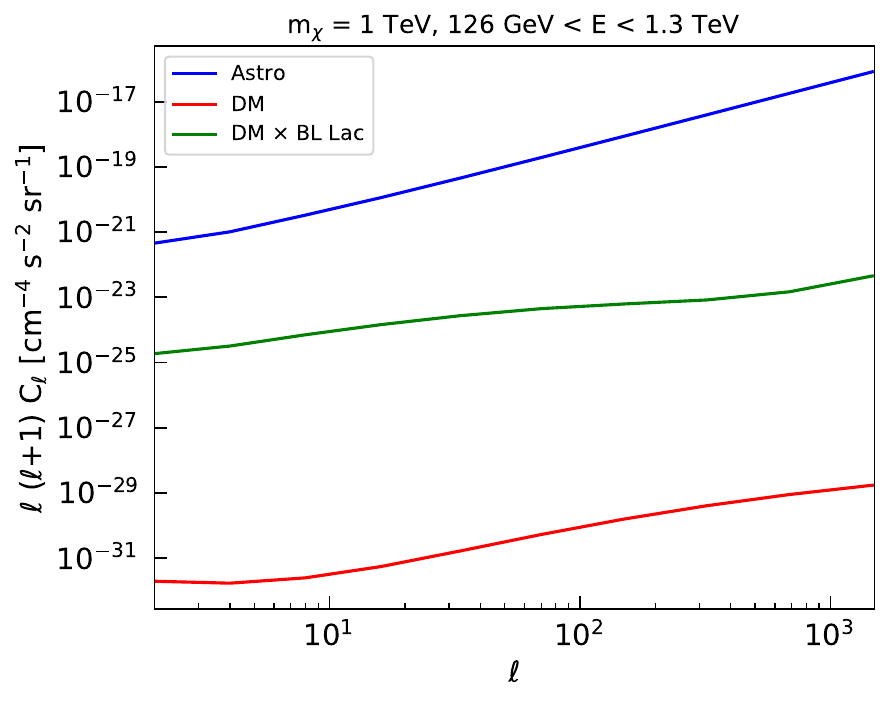}
    \caption{Angular power spectra of the $\gamma$-ray auto-correlation as a function of the multipole, for two different energy bins. The upper (blue) curve refers to the auto-correlation among astrophysical sources, the median (green) and lower (red) curves to the correlation between sources and dark matter and the autocorrelation of dark matter, respectively, for a dark matter mass of 1 TeV and annihilation into $b\bar{b}$ with a cross-section $\langle \sigma v \rangle = 3 \times 10^{-26}$cm$^{-3}$s$^{-1}$.}
    \label{fig:Cl_1TeV_binned}
\end{figure}

\begin{figure}[t!]
    \centering
    \includegraphics[width=0.45\textwidth]{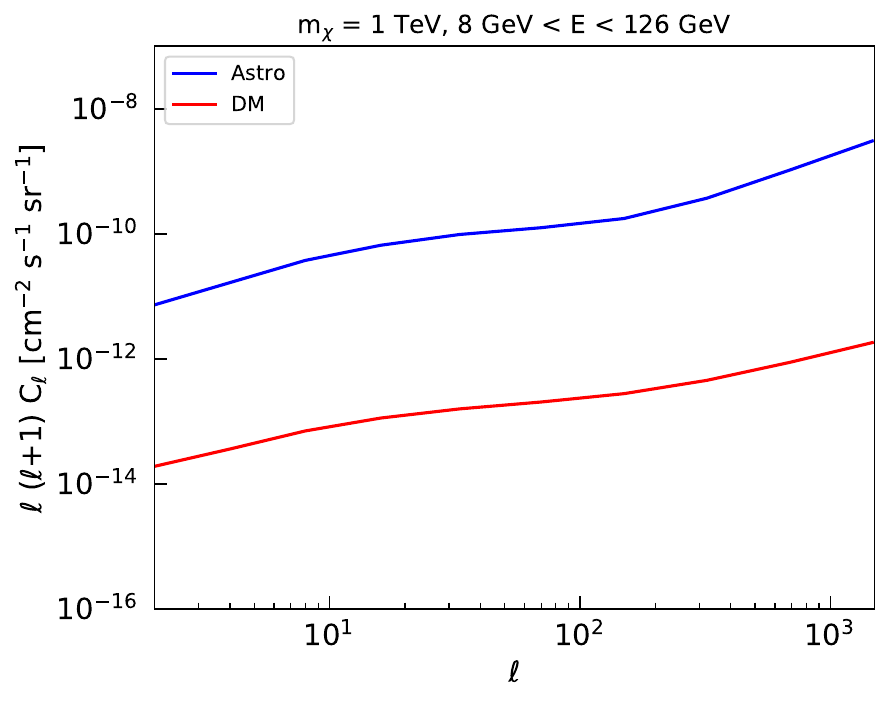}
    \includegraphics[width=0.45\textwidth]{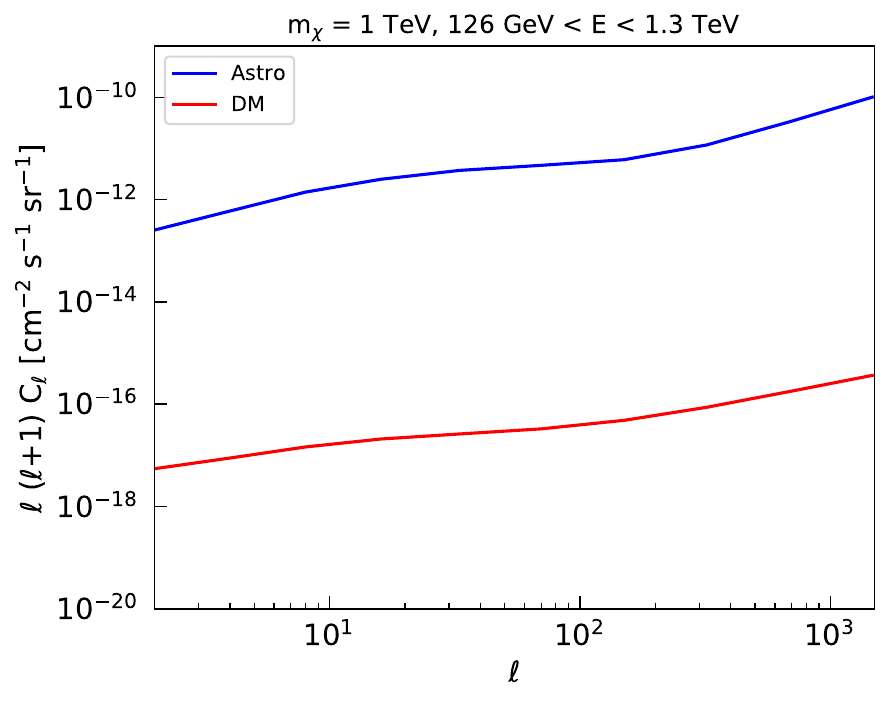}
    \caption{Angular power spectra of the $\gamma$-ray cross-correlation with 2MASS galaxies, as a function of the multipole, for two different energy bins. The upper (blue) curve refers to the cross-correlation with astrophysical sources, while the lower (red) curves represent the correlation with a dark matter signal produced by a particle with mass of 1 TeV and annihilation cross-section into $b\bar{b}$ of $\langle \sigma v \rangle = 3 \times 10^{-26}$cm$^{-3}$s$^{-1}$.}
    \label{fig:Cl_1TeV_cross}
\end{figure}

\begin{figure}[t!]
    \centering
    \includegraphics[width=0.45\textwidth]{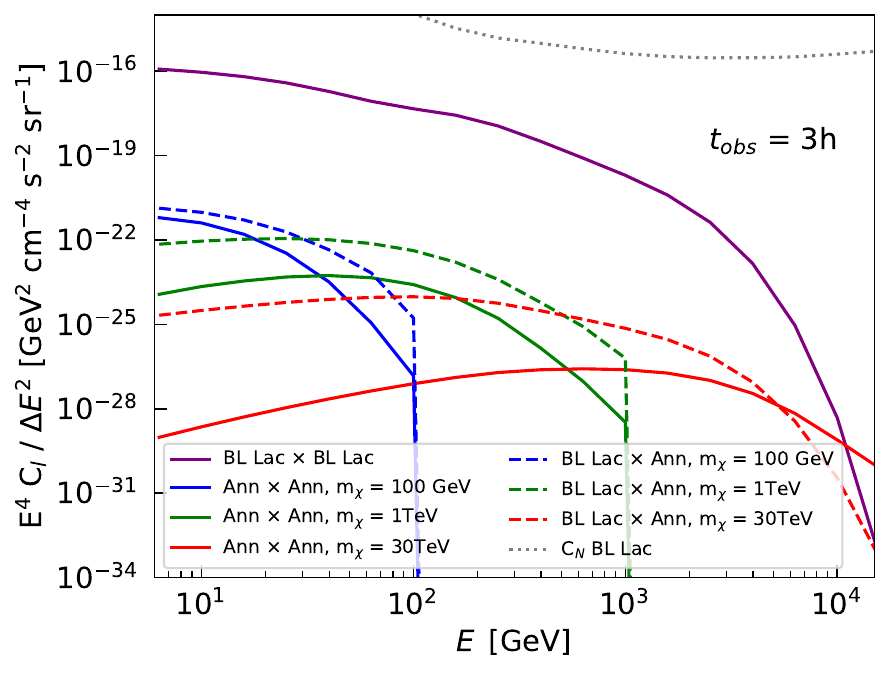}
    \includegraphics[width=0.45\textwidth]{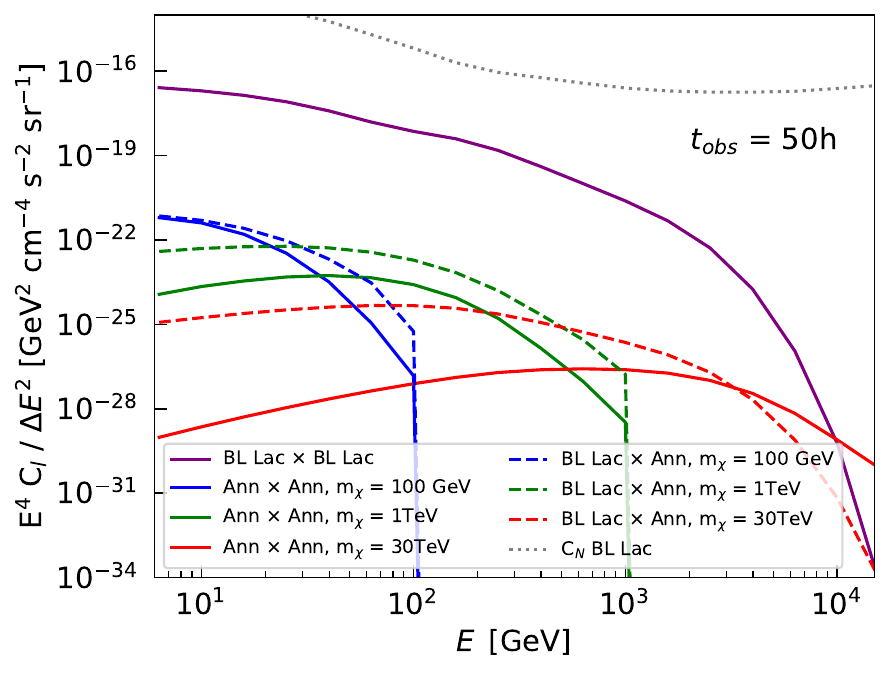}
    \includegraphics[width=0.45\textwidth]{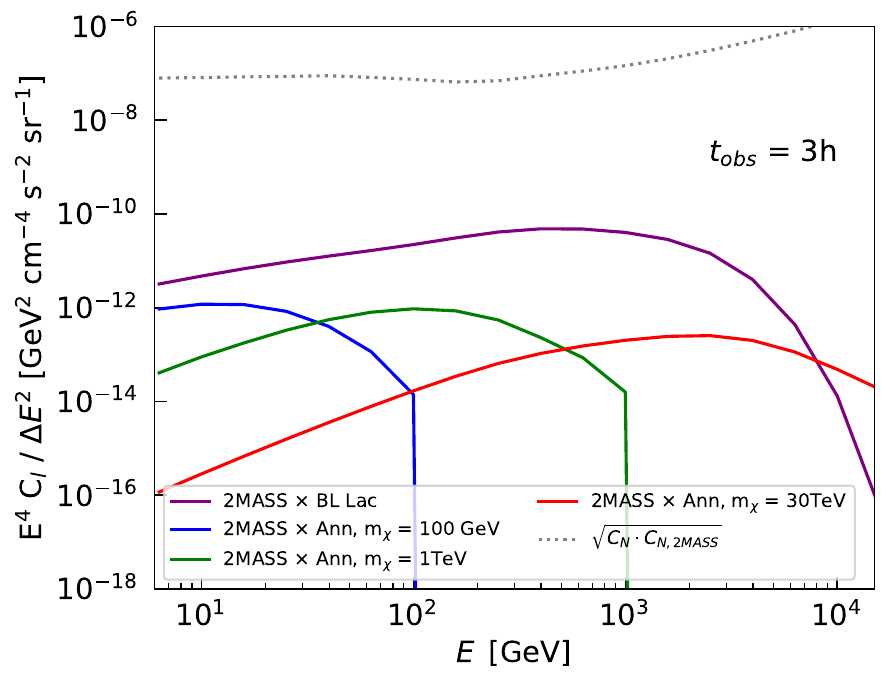}
    \includegraphics[width=0.45\textwidth]{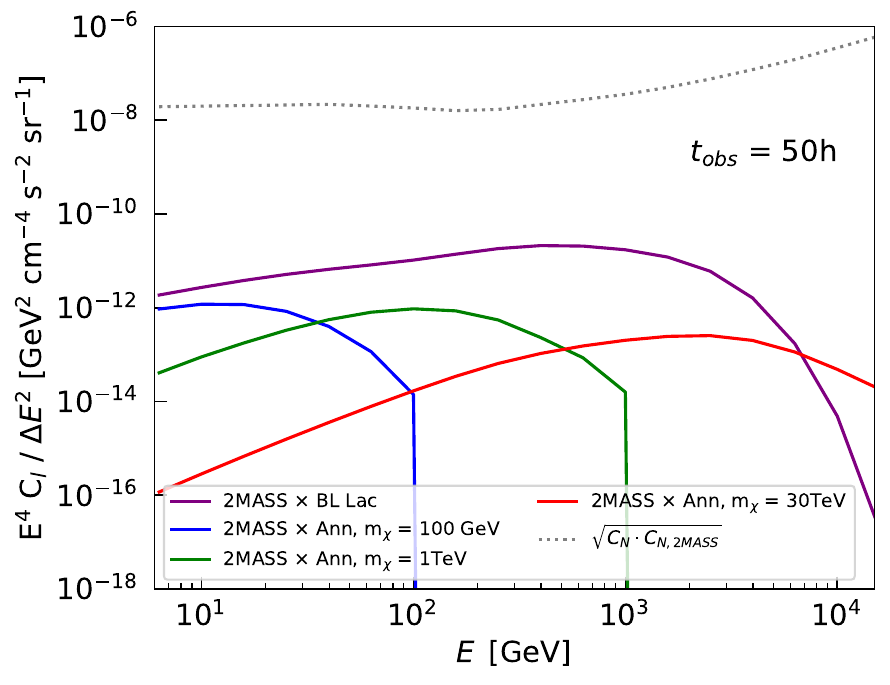}
    \caption{Angular power spectra $C_\ell$ as a function of energy for $\ell=100$. The left panels correspond to 3 hours of observation, while the right panels represent 50 hours. Dark matter signals are calculated for annihilation into $b\bar{b}$, for $\langle \sigma v \rangle = 3 \times 10^{-26}$cm$^{-3}$s$^{-1}$, and three representative masses: 100 GeV, 1 TeV, and 30 TeV. The solid and dashed lines show the auto-correlation and the cross-correlation $C_\ell$ for various combinations of $\gamma$-ray emitters and galaxy catalogs, as labelled. The black dotted lines in the topmost part of the panels represent the noise level for each category: for the upper row ($\gamma$-ray auto-correlations), the noise is dominated by the photon noise of astrophysical sources, and for the lower row (cross-correlation of $\gamma$-ray with galaxies), the geometric mean of the photon and galaxy noise is shown as a representative noise level.} 
    \label{fig:Cl_vs_E}
\end{figure}

Figs. \ref{fig:PS_tot}-\ref{fig:P1h_P2h} report the plots of the various 3D power spectra adopted in the analysis. Figs.\ref{fig:Cl_1TeV_binned}-\ref{fig:Cl_1TeV_cross} show a few samples of auto and cross-correlation APS as a function of the multipole $\ell$. It can be seen that auto APS of astrophysical sources is constant in $\ell$, since they are point-like, while the DM APS presents some structure due to the extension of the DM halos. Fig.\ref{fig:Cl_vs_E}, instead, shows the APS as a function of energy at a fixed multipole $\ell=100$. For comparison, the auto- and cross-correlation noise level is plotted, thus showing that the expected signal-to-noise ratio is typically low.

\subsection{Poisson anisotropy}
\begin{figure}[t!]
    \centering
    \includegraphics[width=0.45\textwidth]{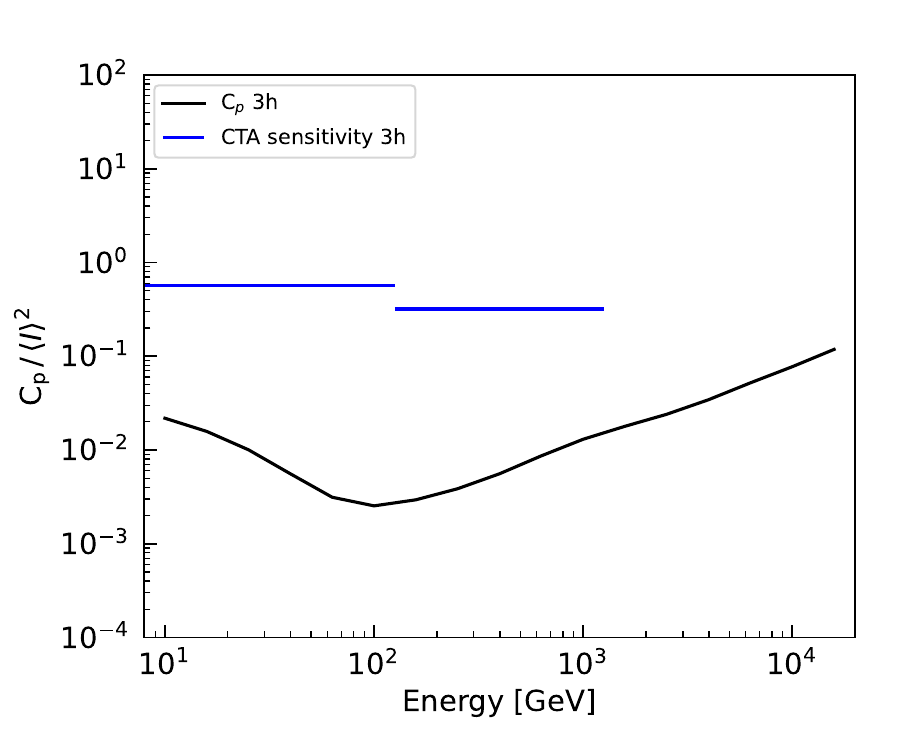}
    \includegraphics[width=0.45\textwidth]{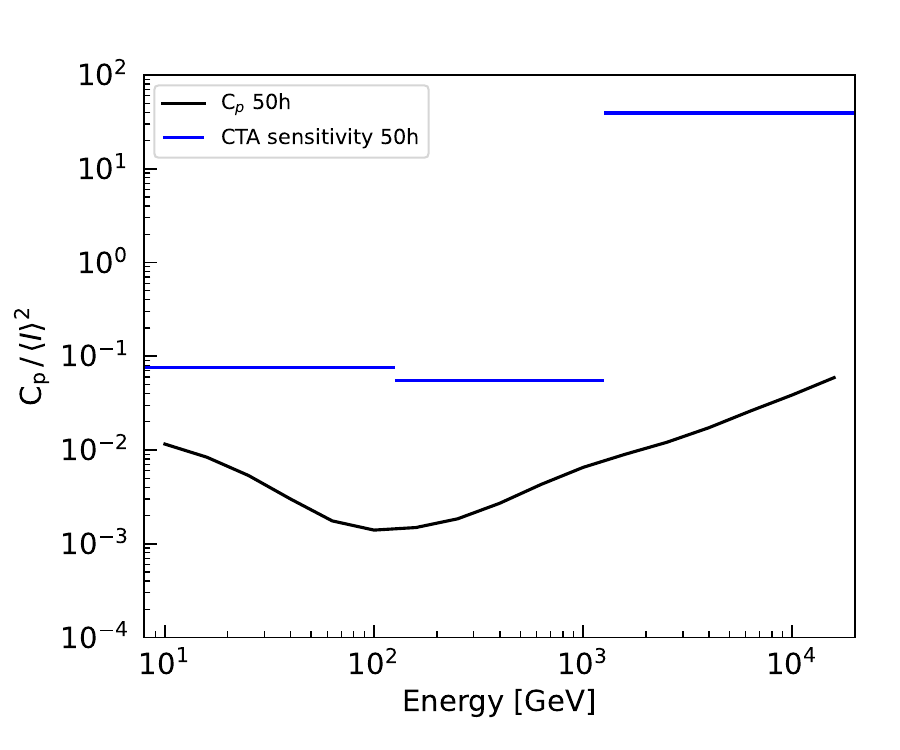}
    \caption{CTA sensitivity to the Poisson anisotropy $C_p$ (normalized to the square of the mean $\gamma$-ray intensity) in the low, mid and high energy bins, for 3 hours (left panel) and 50 hours (right panel) of observation. The predicted sensitivity is to the expected $C_p$ from {\sl unresolved} blazars (solid black lines). The reduced sensitivity at high energies is driven by the limited number of $\gamma$-ray events in this energy range.}
    \label{fig:Cp_I2}
\end{figure}

The auto-correlation angular power spectrum of blazars is dominated by the constant 1-halo term. and is therefore flat in $\ell$. It is thus convenient to define the Poisson anisotropy as the normalization of this spectrum.
\cref{fig:Cp_I2} shows the Poisson anisotropy of unresolved blazars, divided by the square of the intensity as a function of energy for 3 hours and 50 hours of observation. As expected, the anisotropy decreases with increasing observation time, since a larger fraction of the brighter sources (which contribute more significantly to the anisotropy) become resolved. Across the entire energy range, the anisotropy remains at the level of percent or below.

The same figure also shows the CTA sensitivity to the Poisson anisotropy in three energy bins. To derive this sensitivity, we adopt the same methodology used for setting the DM limits. In this case, however, the null hypothesis corresponds to pure noise, that is, the absence of any source contribution, while the alternative hypothesis assumes the presence of a source class characterized by a Poisson anisotropy. We further assume that the energy dependence of the anisotropy of this source class is proportional to $\langle I \rangle^2$, i.e., it scales with the square of the observed intensity.  As shown in the figure, the resulting sensitivity is at the level of approximately 10\%.

\section{Further results}
\label{app:modelll}
\begin{figure}[t!]
    \centering
    \includegraphics[width=0.45\textwidth]{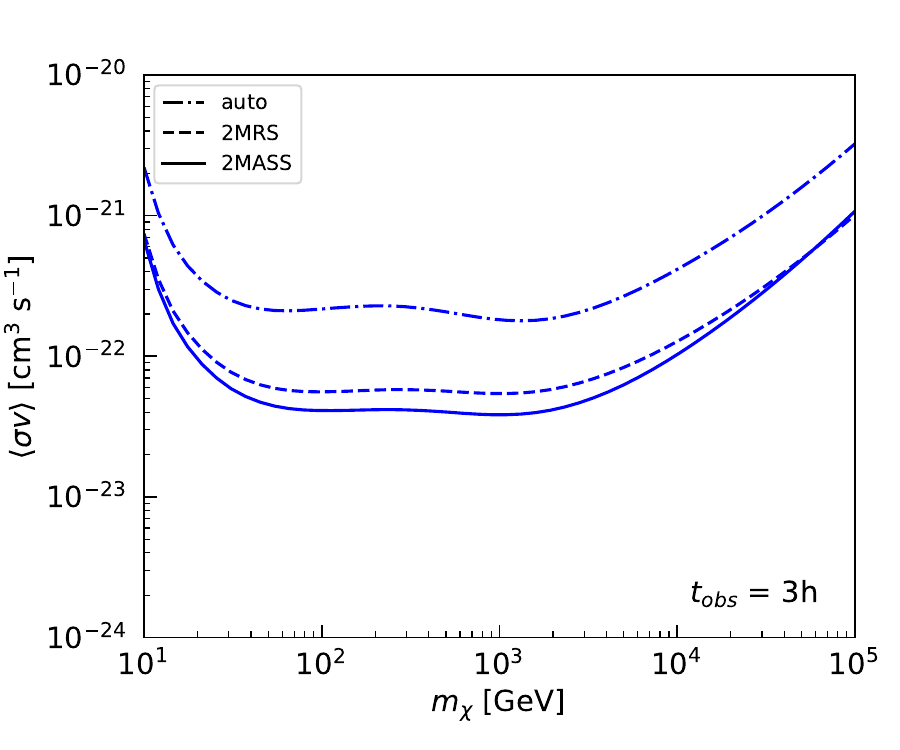}
    \includegraphics[width=0.45\textwidth]{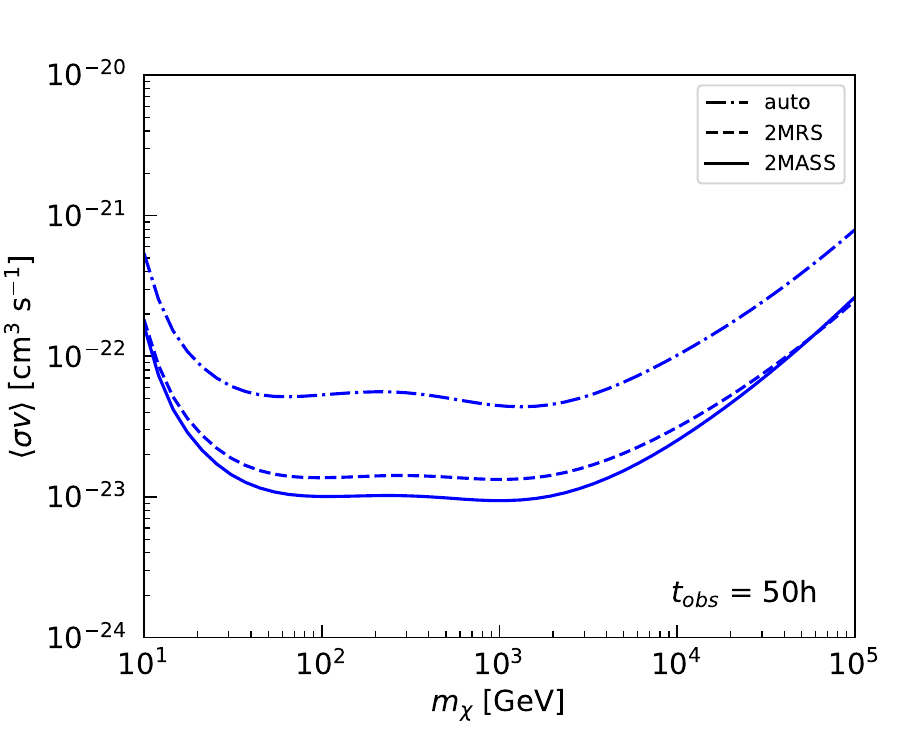}
    \caption{Predicted upper bounds on the dark matter cross-section $\langle \sigma v \rangle$, for annihilation into $b\bar{b}$, for 3 hours (left panel) and 50 hours (right panel) of CTAO observations. The bounds are derived from $\gamma$-ray auto-correlation (dot-dashed), and cross-correlation with 2MRS (dashed lines) and 2MASS (solid) galaxy catalogs.}
    \label{fig:bounds_ann}
\end{figure}

\begin{figure}[t!]
    \centering
    \includegraphics[width=0.45\textwidth]{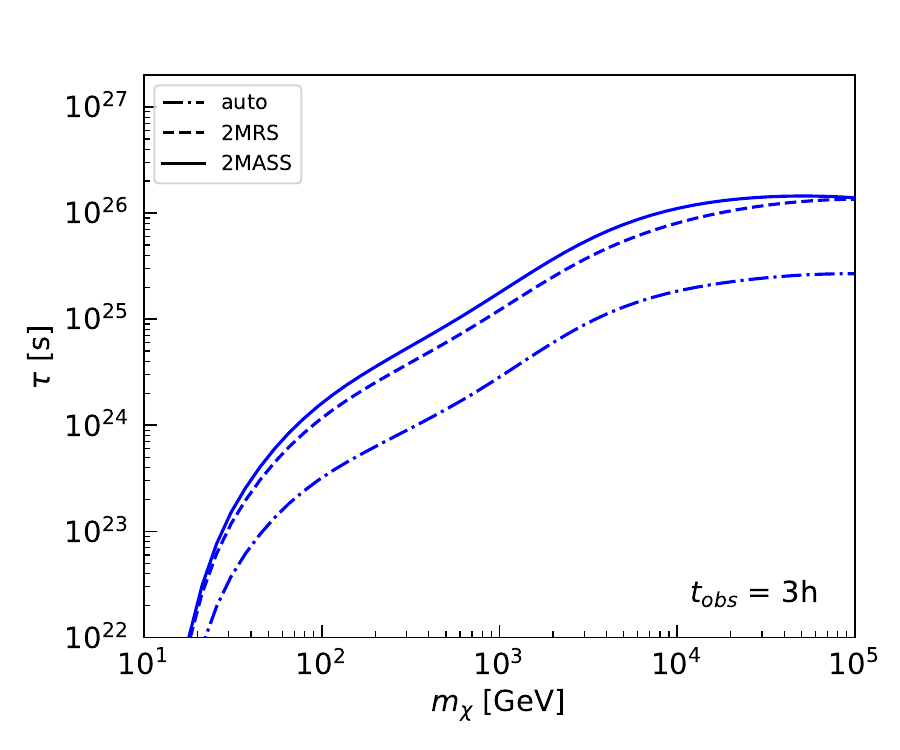}
    \includegraphics[width=0.45\textwidth]{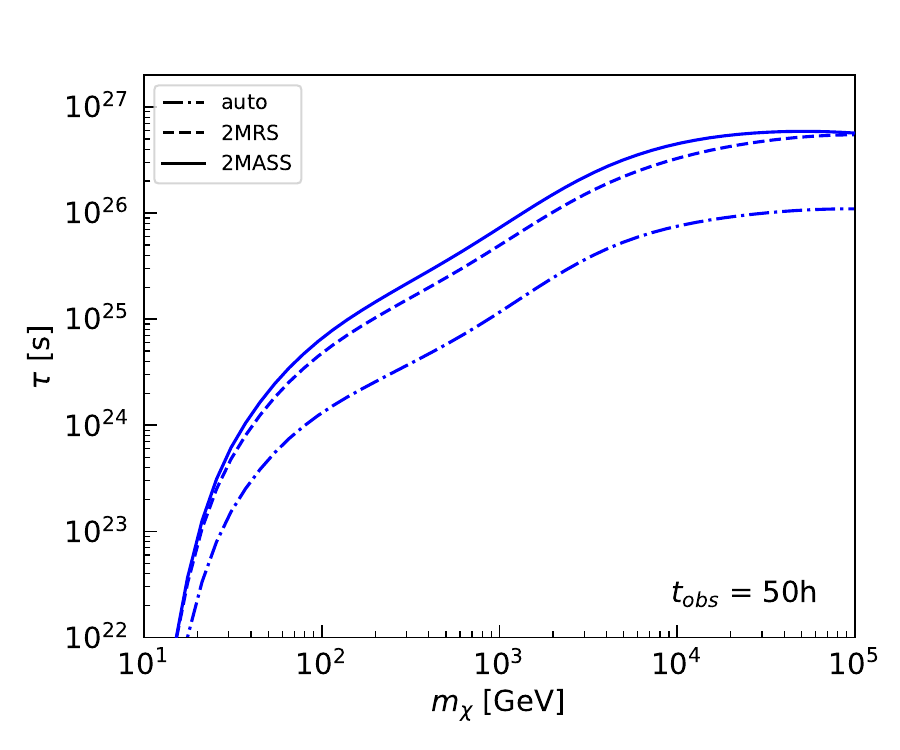}
    \caption{Predicted lower bounds on the dark matter lifetime $\tau$, for decay into $b\bar{b}$, for 3 hours (left panel) and 50 hours (right panel) of CTAO observations. The bounds are derived from $\gamma$-ray auto-correlation (dot-dashed), and cross-correlation with 2MRS (dashed lines) and 2MASS (solid) galaxy catalogs.}
    \label{fig:bounds_decay}
\end{figure}

\begin{figure}[t!]
    \centering
    \includegraphics[width=0.45\textwidth]{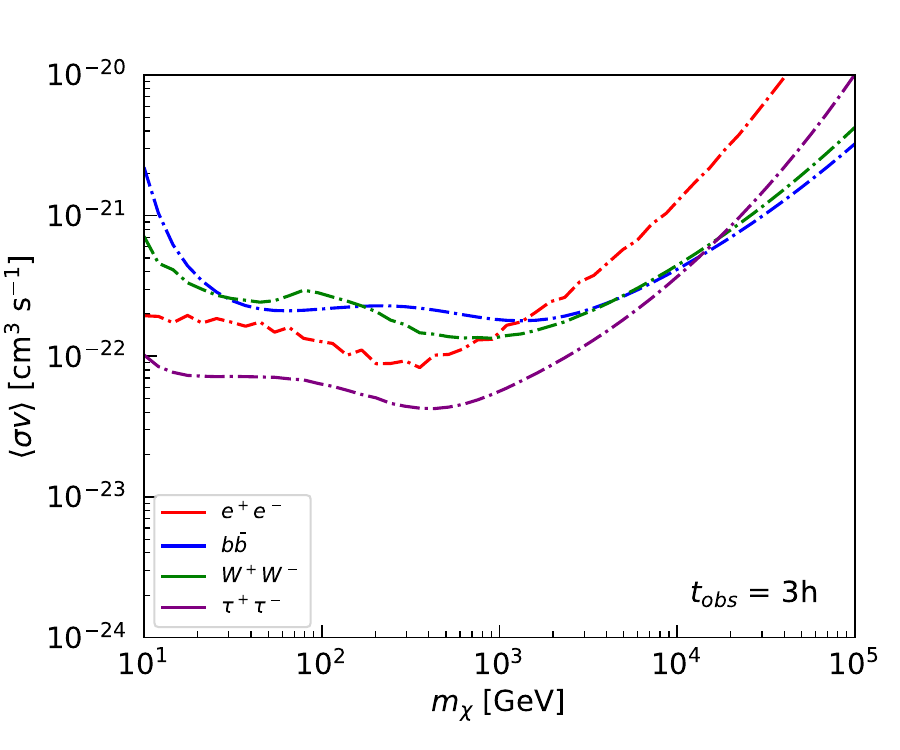}
    \includegraphics[width=0.45\textwidth]{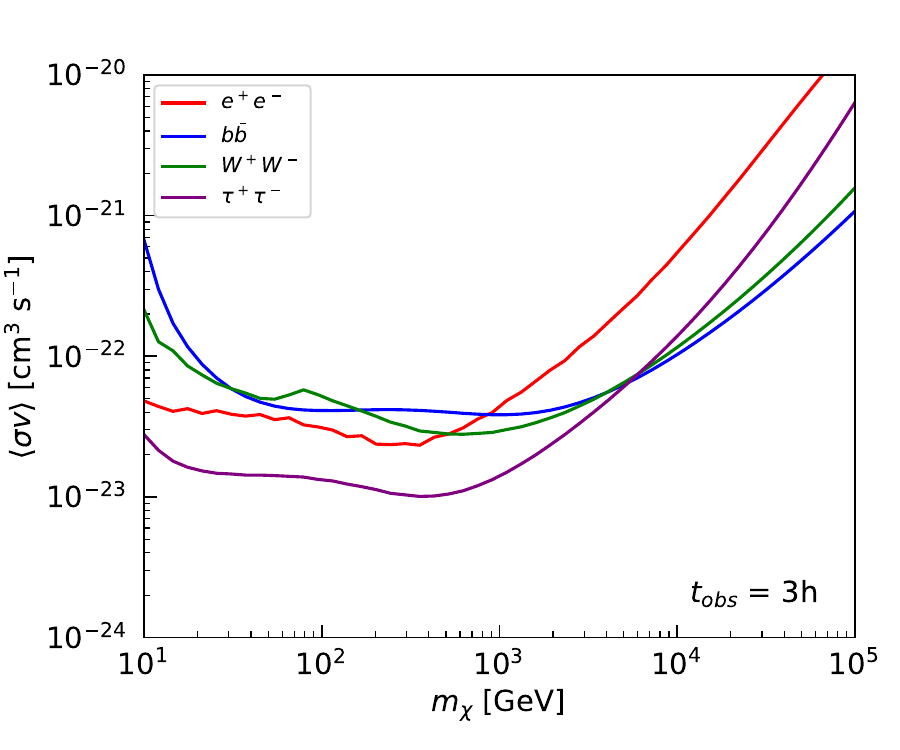}
    \includegraphics[width=0.45\textwidth]{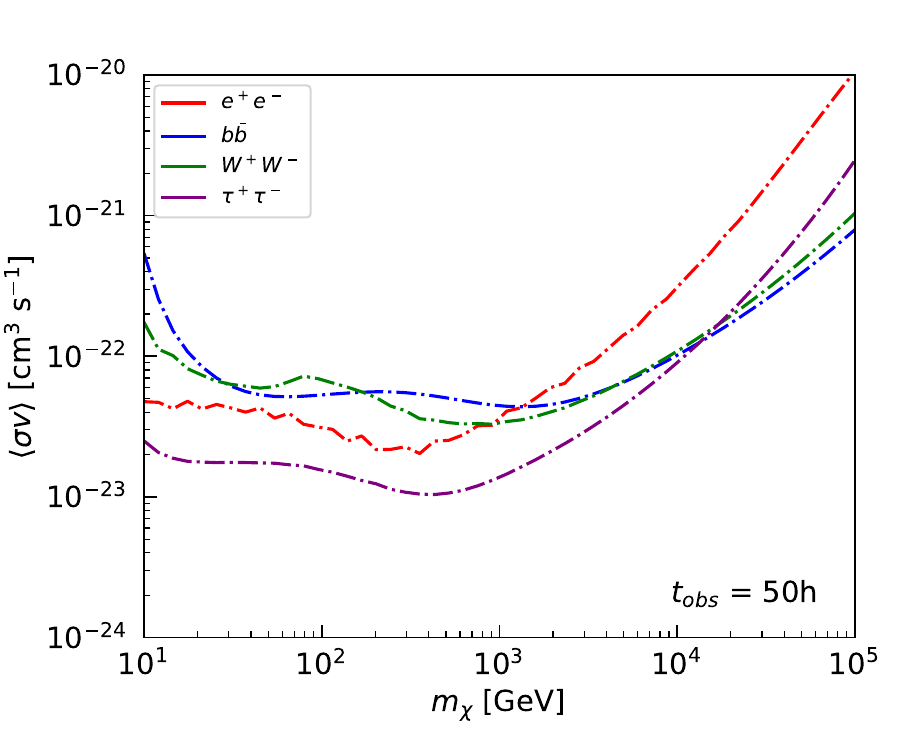}
    \includegraphics[width=0.45\textwidth]{figures/new_analysis/bounds/sigmav_bound_50h_e_b_W_tau_cross.pdf}
    \caption{Predicted upper bounds on the dark matter cross-section $\langle \sigma v \rangle$, for annihilation into various final states: $e^+e^-, b\bar{b}$, $W^+W^-$ and $\tau^+\tau^-$. The upper row shows results for 3 hours of observation; the lower row for 50 hours. Left panels: bounds from $\gamma$-ray auto-correlation. Right panels: bounds from cross-correlation with 2MASS galaxies (the bottom-right panel reproduces the left panel of \cref{fig:all_channels1}).}
    \label{fig:all_channels_ann}
\end{figure}

\begin{figure}[t!]
    \centering
    \includegraphics[width=0.45\textwidth]{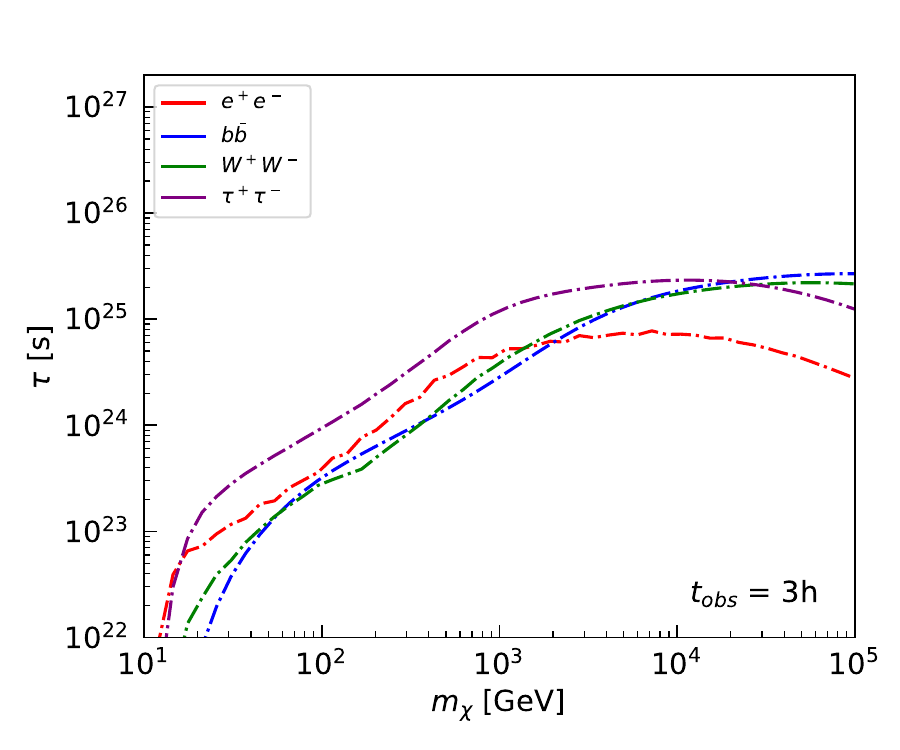}
    \includegraphics[width=0.45\textwidth]{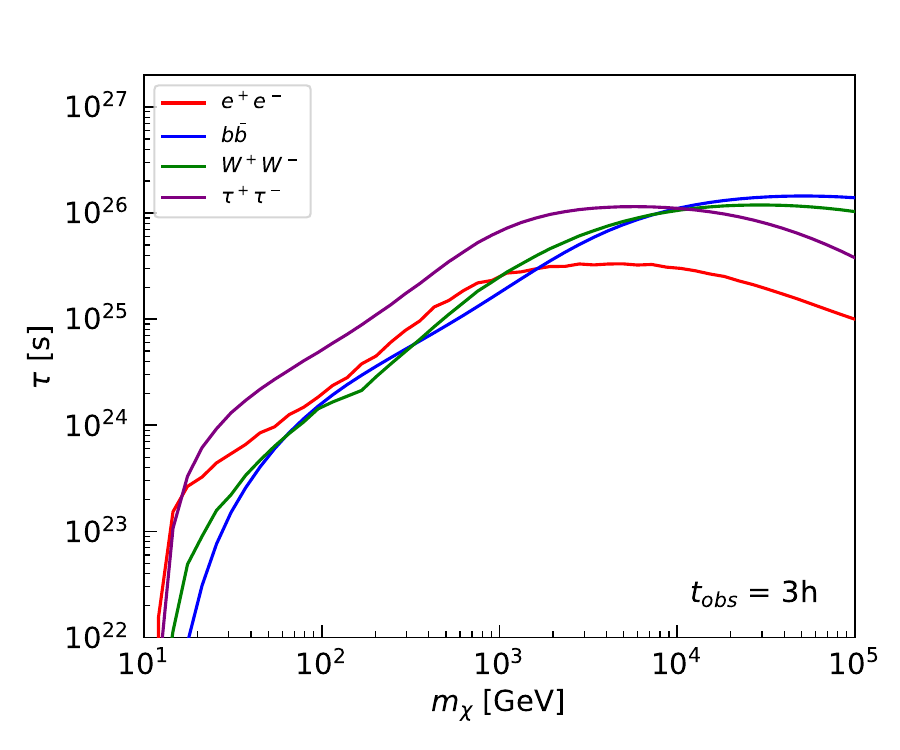}
    \includegraphics[width=0.45\textwidth]{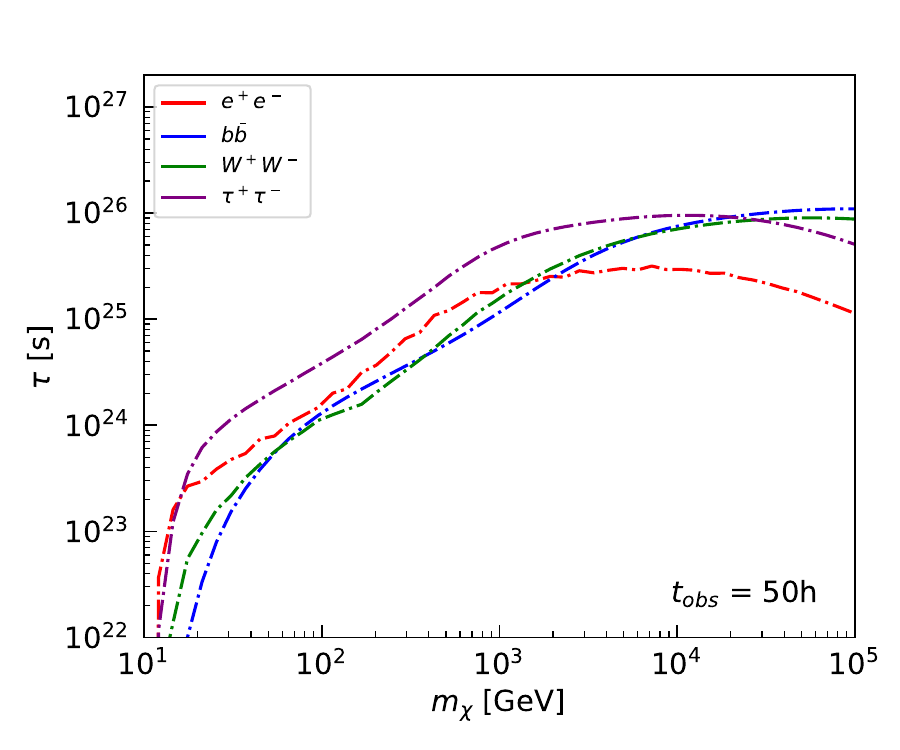}
    \includegraphics[width=0.45\textwidth]{figures/new_analysis/bounds/tau_bound_50h_e_b_W_tau_cross.pdf}
    \caption{Predicted lower bounds on the dark matter lifetime $\tau$, for decay into various final states: $e^+e^-, b\bar{b}$, $W^+W^-$ and $\tau^+\tau^-$. The upper row shows results for 3 hours of observation; the lower row for 50 hours. Left panels: bounds from the $\gamma$-ray auto-correlation. Right panels: bounds from cross-correlation with 2MASS galaxies (the bottom-right panel reproduces the right panel of \cref{fig:all_channels1}).}
    \label{fig:all_channels_dec}
\end{figure}

\begin{figure}[t!]
    \centering
    \includegraphics[width=0.45\textwidth]{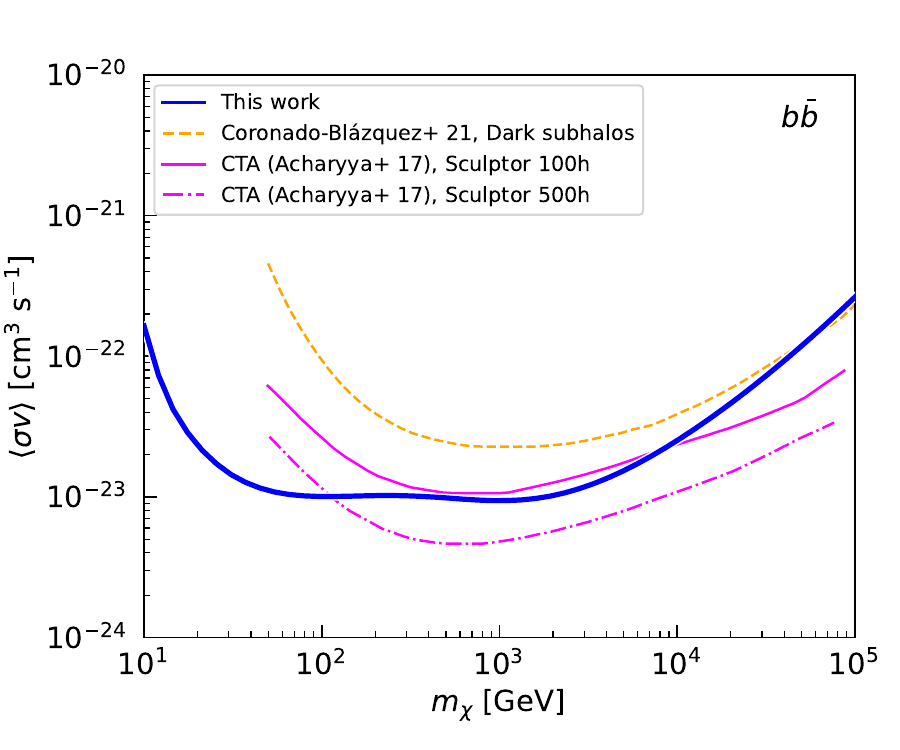}
    \includegraphics[width=0.45\textwidth]{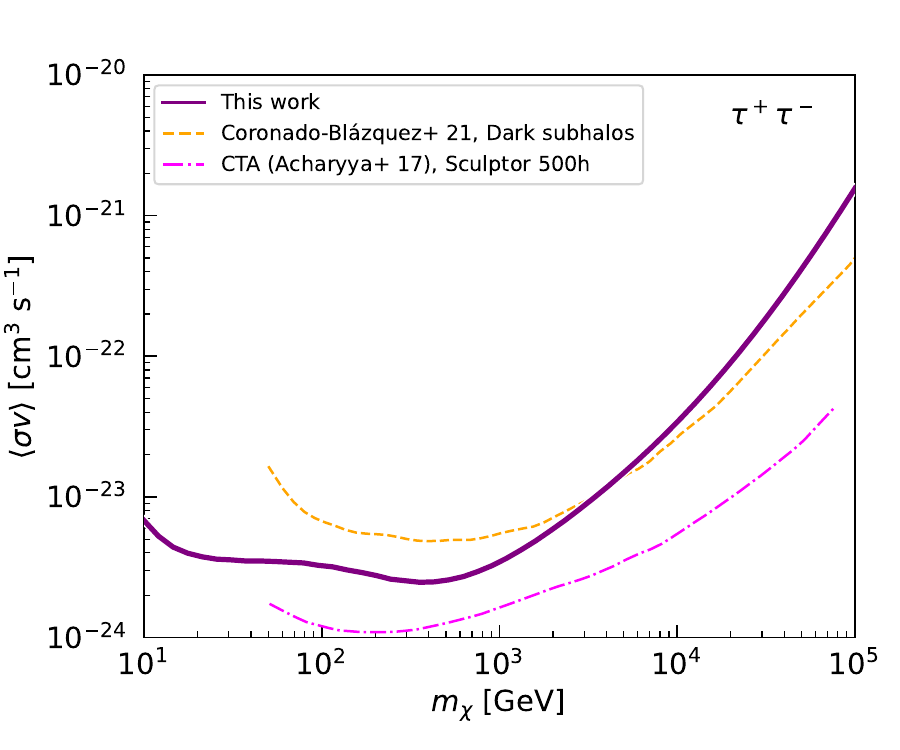}
    \includegraphics[width=0.45\textwidth]{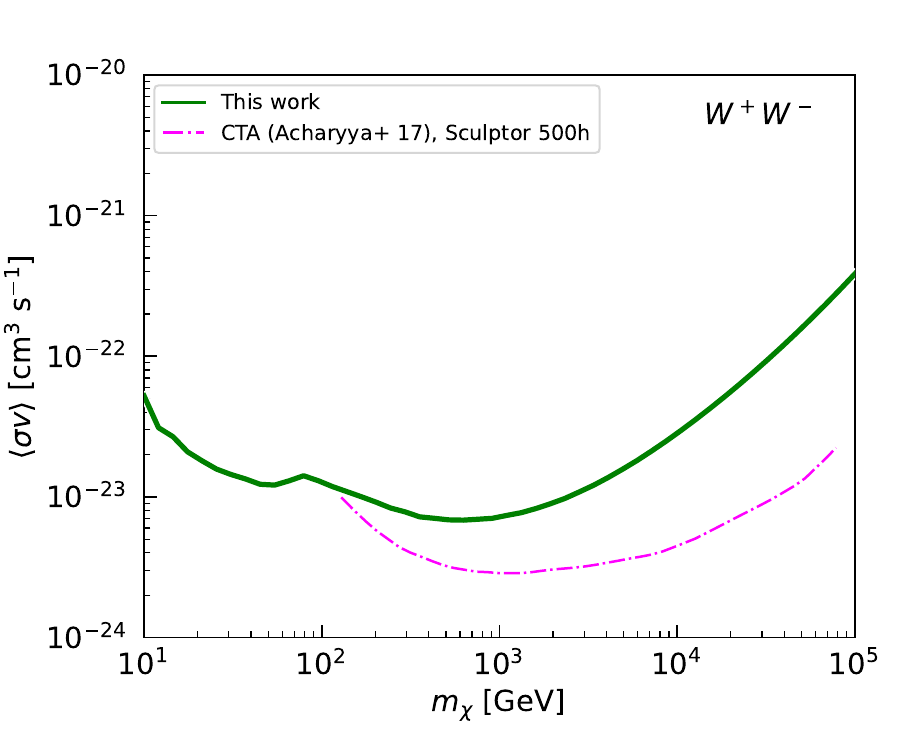}
    \caption{Comparison of the predicted upper bounds on the dark matter annihilation cross-section $\langle\sigma v \rangle$, with existing constraints from the literature: bounds from non-observation of dark sub-halos (dashed yellow) \cite{Coronado_Blazquez_2021}, from the Sculptor dwarf galaxy \cite{CTA_2017} with 100 hours (solid magenta) and 500 hours (dot-dashed magenta) of observations with CTAO. Three channels are displayed: $b\bar{b}$ (upper left panel), $\tau^+\tau^-$ (upper right) and the $W^+W^-$ (lower). The case for the $b\bar{b}$ channel is also shown in \cref{fig:bounds1}.}
    \label{fig:bounds2}
\end{figure}

\begin{figure}[t!]
    \centering
    \includegraphics[width=0.45\textwidth]{figures/new_analysis/bounds/tau_bound_50h_b_cross+lit-bb.pdf}
    \includegraphics[width=0.45\textwidth]{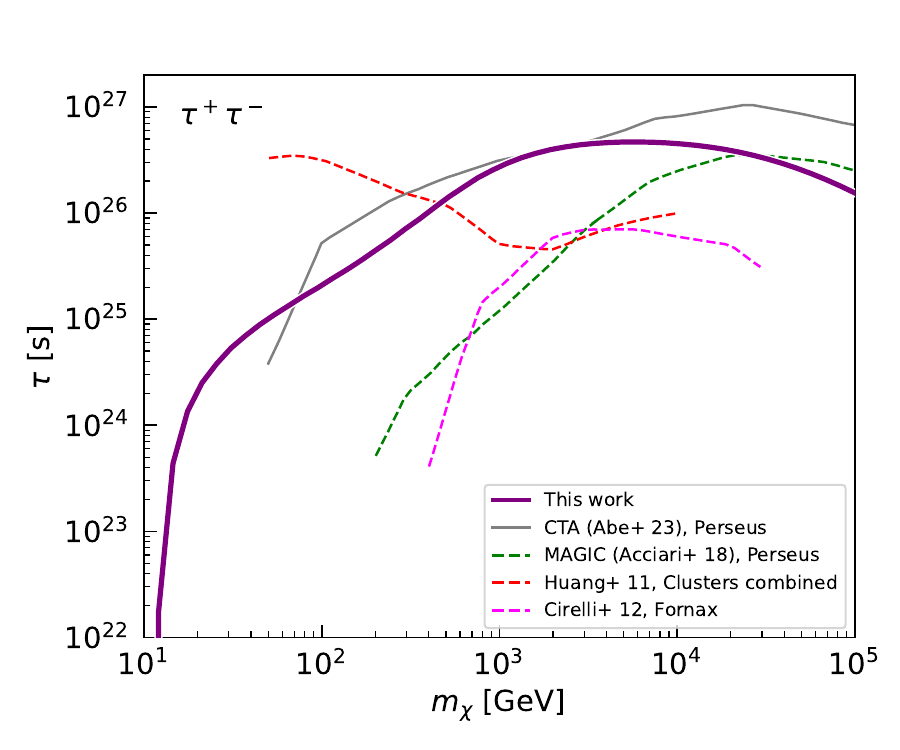}
    \caption{Comparison of the predicted lower bounds on the DM particle lifetime $\tau$ for the $b\bar{b}$ and $\tau^+\tau^-$ decay channels, with existing constraints from the literature: limits from Perseus with CTAO \cite{CTA2023} and MAGIC \cite{MAGIC_2018}, from galaxy clusters \cite{Huang_2012}, and from Fornax \cite{Cirelli_2012}. The case for the $b\bar{b}$ channel is also shown in \cref{fig:bounds1}.}
    \label{fig:bounds3}
\end{figure}

In Figs.~\ref{fig:bounds_ann}, \ref{fig:bounds_decay}, \ref{fig:all_channels_ann}, \ref{fig:all_channels_dec}, \ref{fig:bounds2}, \ref{fig:bounds3}, we show additional results on the constraints for the DM annihilation cross-section and the decay lifetime, exploring various annihilation and decay channels, as well as different observation times. We also provide further comparisons with existing bounds from the literature. 

\end{document}